\pdfoutput=1

\documentclass[11pt,twoside,a4paper,cmspaper,final,collab]{cms-tdr}
\def\svnVersion{130768}\def\cmsCernNo{2011-179}\def\cmsCernDate{\today}\def\cmsMessage{Submitted to the Journal of High Energy Physics}

\begin{document}\cmsNoteHeader{FWD-11-002}

\hyphenation{had-ron-i-za-tion}
\hyphenation{cal-or-i-me-ter}
\hyphenation{de-vices}

\RCS$Revision: 130768 $
\RCS$HeadURL: svn+ssh://alverson@svn.cern.ch/reps/tdr2/papers/FWD-11-002/trunk/FWD-11-002.tex $
\RCS$Id: FWD-11-002.tex 130768 2012-06-19 14:52:04Z alverson $
\newcommand{\DGLAP}{Gribov:1972ri,Lipatov:1974qm,Altarelli:1977zs,Dokshitzer:1977sg}
\newcommand{\BFKL}{Kuraev:1976ge,Kuraev:1977fs,Balitsky:1978ic}
\newcommand{\CCFM}{Ciafaloni:1987ur,Catani:1989yc,Catani:1989sg,Marchesini:1994wr}
\providecommand{\rd}{\ensuremath{\mathrm{d}}}
\cmsNoteHeader{FWD-11-002} 
\title{Measurement of the inclusive  production cross sections for forward jets and for
       dijet events with one forward and one central jet in pp collisions at $\sqrt{s}$~=~7~TeV}

\date{\today}

\abstract{
The inclusive production cross sections for forward jets, as well for jets in dijet events with at least one jet
emitted at central and the other at forward pseudorapidities, are measured
in the range of transverse momenta $\pt =$~35--150\GeVc in proton-proton
collisions at $\sqrt{s}$~=~7~TeV by the CMS experiment at the LHC.
Forward jets are measured within pseudorapidities 3.2~$<|\eta|<$~4.7, and central jets within the $|\eta|<2.8$ range.
The differential cross sections $\rd^2 \sigma /\rd\pt\,\rd\eta$ are compared to
predictions from three approaches in perturbative quantum chromodynamics:
(i) next-to-leading-order calculations obtained with and without
matching to parton-shower Monte Carlo simulations,
(ii) \PYTHIA\ and \HERWIG\ parton-shower event generators with different tunes of parameters,
and (iii) \textsc{cascade} and \textsc{hej} models, including different non-collinear corrections to standard
single-parton radiation.
The single-jet inclusive forward jet spectrum is well described by all models, but not all predictions are consistent
with the spectra observed for the forward-central dijet events.}

\hypersetup{%
pdfauthor={CMS Collaboration},%
pdftitle={Measurement of the inclusive  production cross sections for forward jets and for dijet events with one forward and one central jet in pp collisions at sqrt(s) = 7 TeV},%
pdfsubject={CMS},%
pdfkeywords={CMS, pQCD, jets, forward physics}}

\maketitle 

\section{Introduction}

Jet production in hadron-hadron collisions is sensitive to the nature of the underlying parton-parton scattering processes,
to the details of parton radiation, as well as to the parton distribution functions (PDF) of the colliding hadrons~\cite{Ellis:2007ib}. 
The jet cross sections at large transverse momenta ($\pt$) measured at the Large Hadron Collider
(LHC)~\cite{Chatrchyan:2011me,Aad:2011fc} as well as at previous colliders~\cite{Abazov:2008hua,Aaltonen:2008eq},
are well described over several orders of magnitude by perturbative quantum chromodynamics (pQCD).
However, the measurements are often limited to central pseudorapidities
($|\eta| \lesssim 3$), with $\eta = -\log\left[\tan\left(\theta/2\right)\right]$ where $\theta$ is the polar angle of
the jet with respect to the beam axis. 
In this region of phase space, the momentum fractions $x_1$ and $x_2$ of the incoming partons are of the same magnitude.
Jets emitted at small polar angles ($|\eta| \gtrsim 3$) usually arise from collisions between
partons of significantly different momentum fractions $x_2\ll x_1$, and thereby probe regions of PDF
with contributions from small as well as large $x$ values, which, especially for gluons, are less
well constrained by deep-inelastic scattering data~\cite{HERAPDF10:2009wt}. In the phase
space considered in this paper, gluons participate in about 80\% of the partonic interactions that lead to
forward jet production, with paired parton momentum fractions of
the order of $x_2\approx$~10$^{-4}$ and $x_1\approx$~0.2~\cite{Cerci:2008xv}.

The type of dijet final states studied in this analysis also provides information on multi-parton
production processes with large separations in pseudorapidity whose theoretical description involves multiple
scales and possibly large logarithmic contributions. Such event topologies may show deviations from the parton
radiation patterns expected from the standard Dok\-shit\-zer-Gribov-Lipatov-Altarelli-Parisi (DGLAP) evolution equations~\cite{\DGLAP},
as modelled in the approaches of e.g. Balitski-Fadin-Kuraev-Lipatov (BFKL)~\cite{\BFKL},
Ciafaloni-Catani-Fiorani-Mar\-che\-si\-ni (CCFM)~\cite{\CCFM}, or gluon saturation~\cite{Gelis:2010nm}.
Understanding the dynamics of forward jet production, either with or without accompanying central jets,
is also essential for modelling multijet backgrounds at the LHC, e.g. in Higgs boson searches
in channels involving vector-boson fusion~\cite{Figy:2003nv} or requiring a central-jet veto~\cite{Berger:2010xi},
as well as in extracting vector-boson scattering cross sections~\cite{Butterworth:2002tt}.

The Compact Muon Solenoid (CMS) detector provides a calorimetric coverage to study jet production over a range of
jet pseudorapidities as large as $\Delta\eta \approx$~10 which has not been reached heretofore. The study
presented here considers the measurement of central and forward jets with maximum rapidity separations of $\Delta \eta \approx$~6
similar to a recent ATLAS study~\cite{Aad:2011jz}. 
Here, the inclusive production of forward jets, as well as that of forward jets produced in conjunction with central jets,
is studied in data collected with the CMS detector at the LHC during 2010 in proton-proton (pp) collisions at a centre-of-mass energy of 7 TeV.
The distributions of interest include the single-jet inclusive differential cross section
${\rd^2\sigma}/{\rd\pt\,\rd\eta}$ for forward jets, as well as the differential cross sections
${\rd^2\sigma}/{\rd\pt^{f}\,\rd\eta^f}$ and ${\rd^2\sigma}/{\rd\pt^{c}\,\rd\eta^c}$
for the simultaneous production of at least one forward jet ($f$)
in conjunction with at least one central jet ($c$). The axis of the forward jet is required to be
in the fiducial acceptance of the hadron forward calorimeters (3.2~$<\left|\eta\right|<$~4.7), and
that of the central jet within $\left|\eta\right|<$~2.8. The concurrent measurement of at least one jet
in both $\eta$ regions is referred to as ``dijet'' in the following.

The final jet spectra are fully corrected to the level of stable particles (lifetime $\tau$ with $c\tau >$~10~mm)
coming out from the proton-proton interaction (which we will refer to as ``particle-level'' hereafter),
and compared with predictions from several Monte Carlo (MC) event generators,
such as \PYTHIA~6 (version 6.422)~\cite{pythia}, \PYTHIA~8 (version 8.135)~\cite{Sjostrand:2007gs},
\HERWIG~6 (version 6.510.3)~\cite{herwig6} + \textsc{Jimmy}~\cite{jimmy}, and \textsc{herwig++} (version 2.3)~\cite{Bahr:2008tx}.
The data are also compared to next-to-leading-order (NLO) pQCD predictions obtained either
with \textsc{nlojet++}~\cite{Nagy:2001fj,Nagy:2003tz} corrected for non-perturbative effects,
or with the \POWHEG\ package~\cite{Frixione:2007vw} which implements a matching to \PYTHIA\ or \HERWIG\ parton showers.
In addition, the measured distributions are compared to results from the \textsc{cascade} (version 2.2.04)~\cite{Jung:2000hk,Jung:2010si}
and \textsc{hej}~\cite{Andersen:2009nu,Andersen:2011hs} programs. \textsc{cascade} includes parton radiation from QCD evolution
in $1/x$ and \textsc{hej} includes extra contributions from wide-angle gluon radiation, that are not provided in the other models.

This paper is organised as follows. Sections~\ref{sec:Detector} and~\ref{sec:sample_evt_select}
describe the experimental apparatus and the data sample used in the analysis.
Jet reconstruction and energy corrections are presented in Sections~\ref{sec:jetreco}
and~\ref{sec:jetcorr}, respectively.
The results and their associated uncertainties, discussed in Section~\ref{sec:spectra},
are compared to theoretical expectations in Section~\ref{sec:data_vs_theory}, and
the conclusions are summarised in Section~\ref{sec:Conclusions}.

\section{Experimental setup}
\label{sec:Detector}

The CMS detector is described in Ref.~\cite{Adolphi:2008zzk}. Only the detector systems used in this analysis
are discussed hereafter. The central feature of the CMS detector is a superconducting solenoid
that provides an axial magnetic field of 3.8\unit{T} parallel to the beam axis.
Charged particle trajectories are measured using silicon pixels and strip trackers that cover the
pseudorapidity region $|\eta|<$~2.5.
An electromagnetic crystal calorimeter (ECAL) and a brass/scintillator hadron
calorimeter (HCAL) surround the tracking volume and cover
$|\eta|<$~3.0. A forward quartz-fibre Cherenkov hadron calorimeter (HF)
extends the coverage to $|\eta|$~=~5.2. 

The relevant detector components for the reconstruction of jets in this work are the ECAL and HCAL
central calorimeters~\cite{Chatrchyan:2009qm,Chatrchyan:2009vn}, as well as the HF forward calorimeters~\cite{HF}.
The ECAL cells are grouped in quasi-projective towers of granularity in pseudorapidity and azimuthal
angle of $\Delta \eta \times \Delta\phi$~=~0.0174$\times$0.0174 in the barrel ($|\eta| <$~1.5),
and of 0.05$\times$0.05 in the endcap (1.5~$<|\eta|<$~3.0).
The HCAL has a tower granularity as small as $\Delta \eta \times \Delta\phi =$~0.087$\times$0.087.
The HCAL, when combined with the ECAL, measures jets with a resolution $\Delta E/E \approx 100\,\% /\sqrt{E\,(\GeVns)} \oplus 5\,\%$.
The HF calorimeters consist of steel absorbers containing embedded radiation-hard quartz fibres,
located at $\pm 11.2\unit{m}$ from the centre of the CMS detector, and cover the region 2.9~$<|\eta|<$~5.2.
Half of the fibres run over the full longitudinal depth of the absorber, while the
other half start at a depth of 22\unit{cm} from the front face of each detector.
The segmentation of the HF calorimeters is $0.175\times0.175$, except for $|\eta|>$~4.7, where it is $0.175\times 0.35$.
The HF energy resolution is $\sim$200\%/$\sqrt{E(\GeVns)}$.
\section{Data selection}
\label{sec:sample_evt_select}

For online selection, CMS uses a two-level trigger system consisting of a Level-1 and a High Level Trigger (HLT).
The HLT searches for jets using an iterative cone algorithm ~\cite{Bayatian:2006zz,CMS-PAS-JME-07-003} of radius
$R = \sqrt{(\Delta \eta)^2 + (\Delta \phi)^2} = 0.5$.
Events for the inclusive forward-jet analysis were selected with a trigger requiring a minimum jet transverse energy of 15\GeV
within $|\eta|<$~5.2, while the events used in the dijet measurement were taken with a dijet trigger requiring
two jets with summed calorimeter transverse energy above 30\GeV also within $|\eta|<$~5.2.
The total data sample collected at luminosities of about $10^{30}\percms$
corresponds to an integrated luminosity of (3.14~$\pm$~0.14)\pbinv.
Trigger efficiencies are determined from the ratio of the yield of events containing either forward or
forward-central jets that pass the HLT requirements over the yield of events that
pass the minimum-bias and low-threshold (6\GeV) jet-monitor triggers.
In all cases, the HLT is fully efficient 
for single jets with calibrated $\pt >$~35\GeVc.

All events are required to have a primary vertex reconstructed from at least 5 tracks,
consistent with the known transverse position of the beams and within $\pm$24\unit{cm}
of the centre of the detector along the longitudinal direction.
Events are further filtered out in the pixel detector, by requiring
more than 25\% well-reconstructed tracks in events with 10 or more tracks~\cite{track}.
Events with anomalous noise in HF calorimeters, e.g. due to energetic charged particles
that interact in the window of the HF photo-multipliers,
are flagged with different algorithms and rejected. 
These selection criteria reject non-collision and beam-related backgrounds and are highly efficient
($\sim$100\%) for the final states in this analysis.

\section{Jet reconstruction}
\label{sec:jetreco}

The anti-$k_T$ jet clustering algorithm~\cite{a-ktalg,Cacciari:2011ma} is used to reconstruct forward and
central jets with a distance parameter $R$~=~0.5.
The inputs to the clustering correspond to depositions of energies in calorimeter cells and their angles relative to the beam axis.
A four-momentum is associated to each jet by summing the energy of the cells above a given threshold, assuming  zero mass for each cell deposit,
with momentum components specified by the angles of each cell relative to the point of interaction given by the main event vertex~\cite{Chatrchyan:2011ds}.
In the central region, jets are obtained from signals in calorimeter towers with energies in at least one HCAL
cell, and from their geometrically corresponding ECAL crystals.
In the forward region, jets are reconstructed using Cherenkov-light signals collected in both the HF short and long quartz readout fibres.

The central and forward jet regions are defined, respectively, as $|\eta|<$~2.8 and 3.2~$<|\eta|<$~4.7,
where $\eta$ corresponds to the reconstructed jet axis vector applied on the interaction point.
Both $|\eta|$ ranges guarantee full jet reconstruction within the maximum calorimeters limits
taking into account the jet size parameter of $R$~=~0.5.
All jets are required to have a transverse momentum above $\pt=$~35\GeVc.
If more than one jet is present in either the central or forward region, the one with highest $\pt$ is considered in this analysis.
Central jets are required to satisfy the calorimeter quality criteria of Ref.~\cite{Chatrchyan:2011ds}.
In the HF calorimeter, the applied jet quality selections remove unphysical energy depositions. 
These criteria require each jet to have a minimum ($\pt$-dependent) number of HF cells clustered into a jet, and
the fraction of the electromagnetic to total jet energies to be above a parameterised threshold.

\section{Jet energy corrections}
\label{sec:jetcorr}

The jet $\pt$ spectra reconstructed from the calorimeter energies are corrected to account for the following
systematic effects: (i) $\pt$- and $\eta$-dependent response of the calorimeters, and possible overlap with other proton-proton interactions (pileup),
and (ii) an ``unfolding'' of the impact of the jet energy resolution on the migration of events across $\pt$ bins,
and thereby correct the measured spectrum to the particle-level through comparison with MC events, as discussed below.

The reconstructed jet energy scale (JES) is first calibrated
using data based on balancing the $\pt$ values in dijet and in photon-jet events,
as well as from MC simulations~\cite{Chatrchyan:2011ds}.
The ensuing JES corrections adjust the energies according to the relative $\eta$ and $\pt$ dependencies
of the response of the ECAL, HCAL and HF calorimeters. These corrections, with values from 1.0 to 1.2 for HF,
set the absolute energies to their calibrated JES values and also account for the extra pileup energy. 
The latter effect is very small in this analysis which is
mostly based on data collected with un-prescaled low-$\pt$ jet triggers during the early running
of the LHC with an average of $\sim$2.2 collisions per colliding pair of proton bunches.

\begin{figure}[!Hhtb]
\centering
\subfigure[]{\includegraphics[angle=0.,width=0.333\textwidth,height=6.1cm,clip=true]{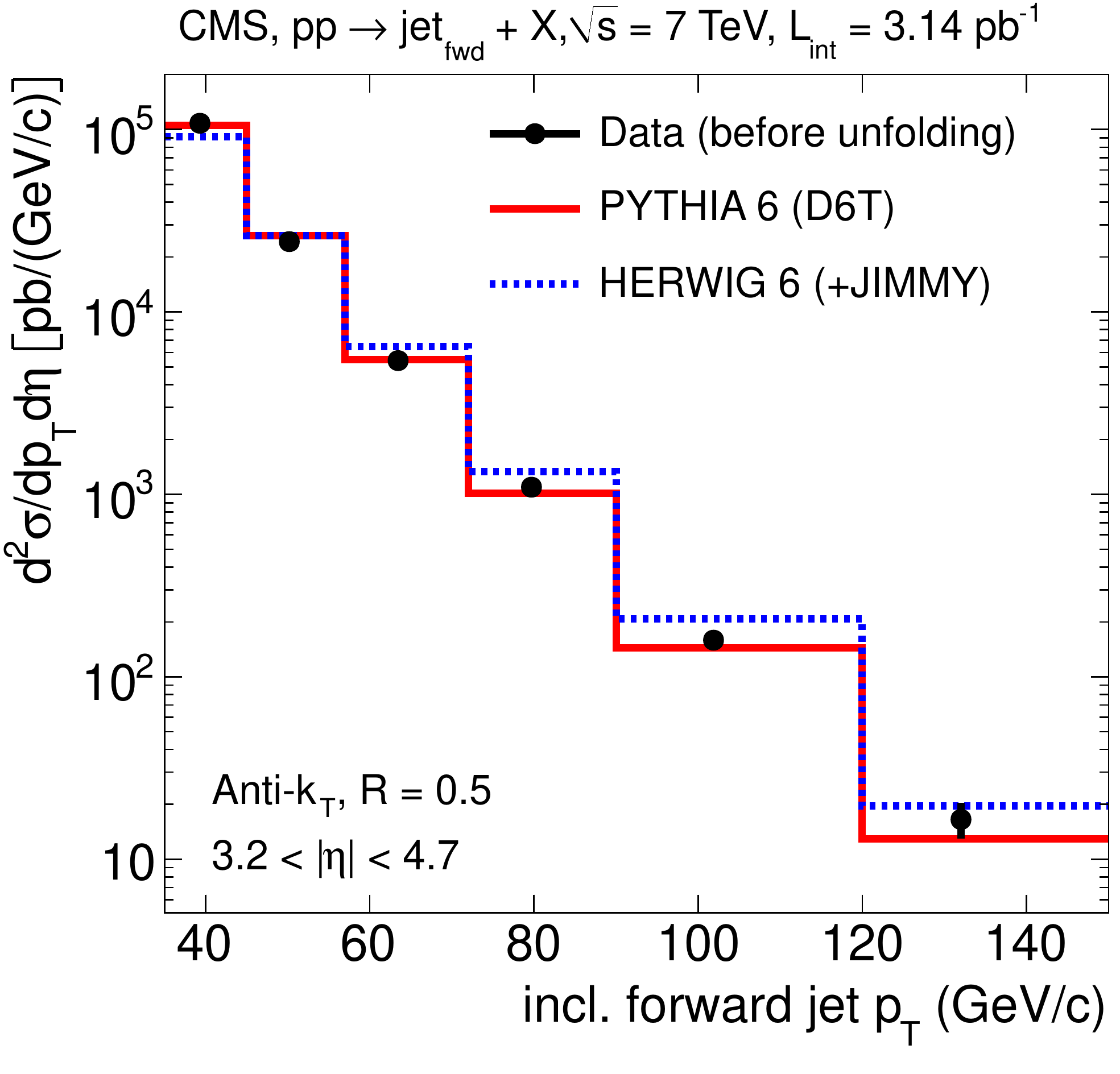}}
\subfigure[]{\includegraphics[angle=0.,width=0.333\textwidth,height=6.1cm,clip=true]{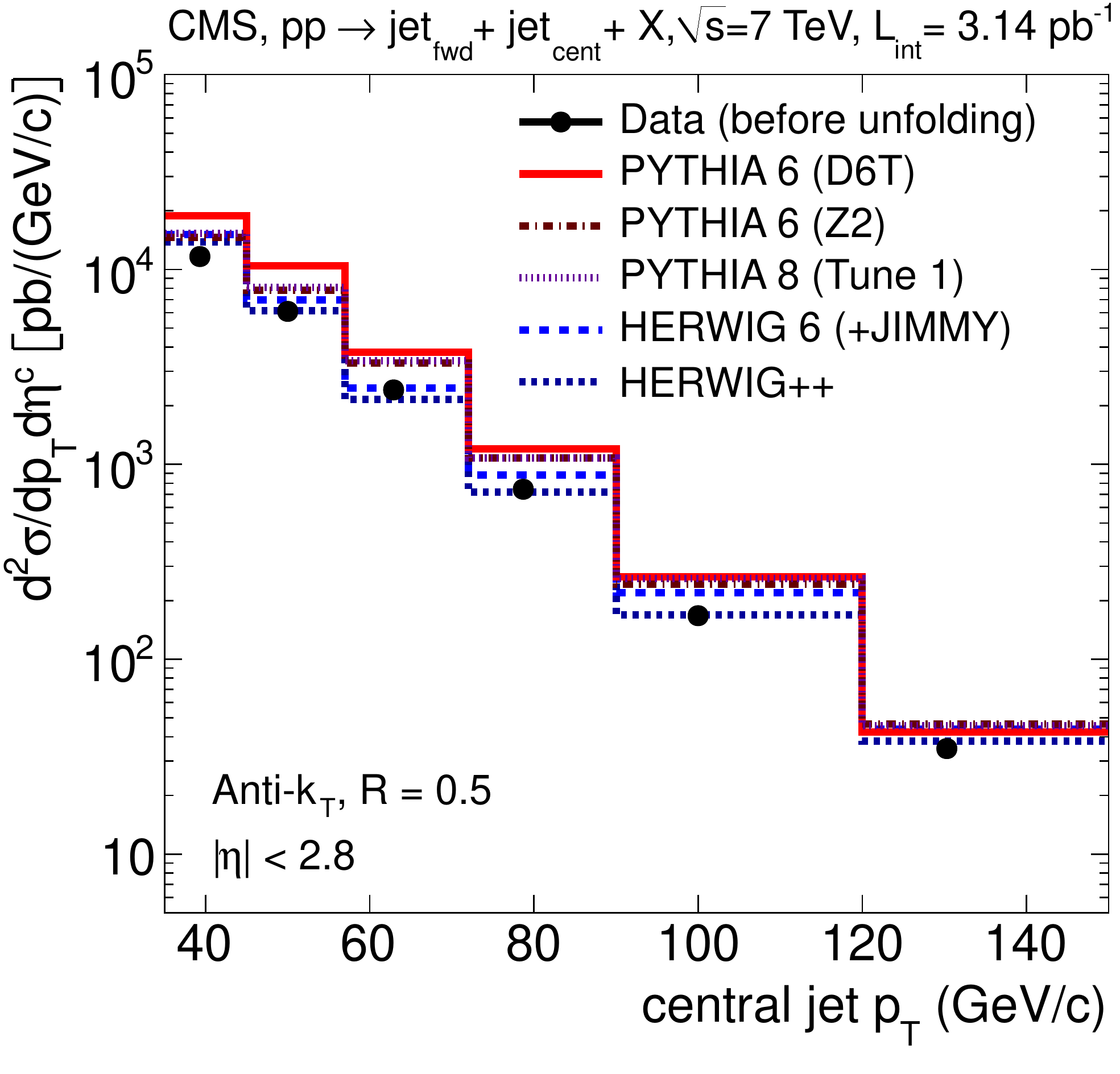}}
\subfigure[]{\includegraphics[angle=0.,width=0.333\textwidth,height=6.1cm,clip=true]{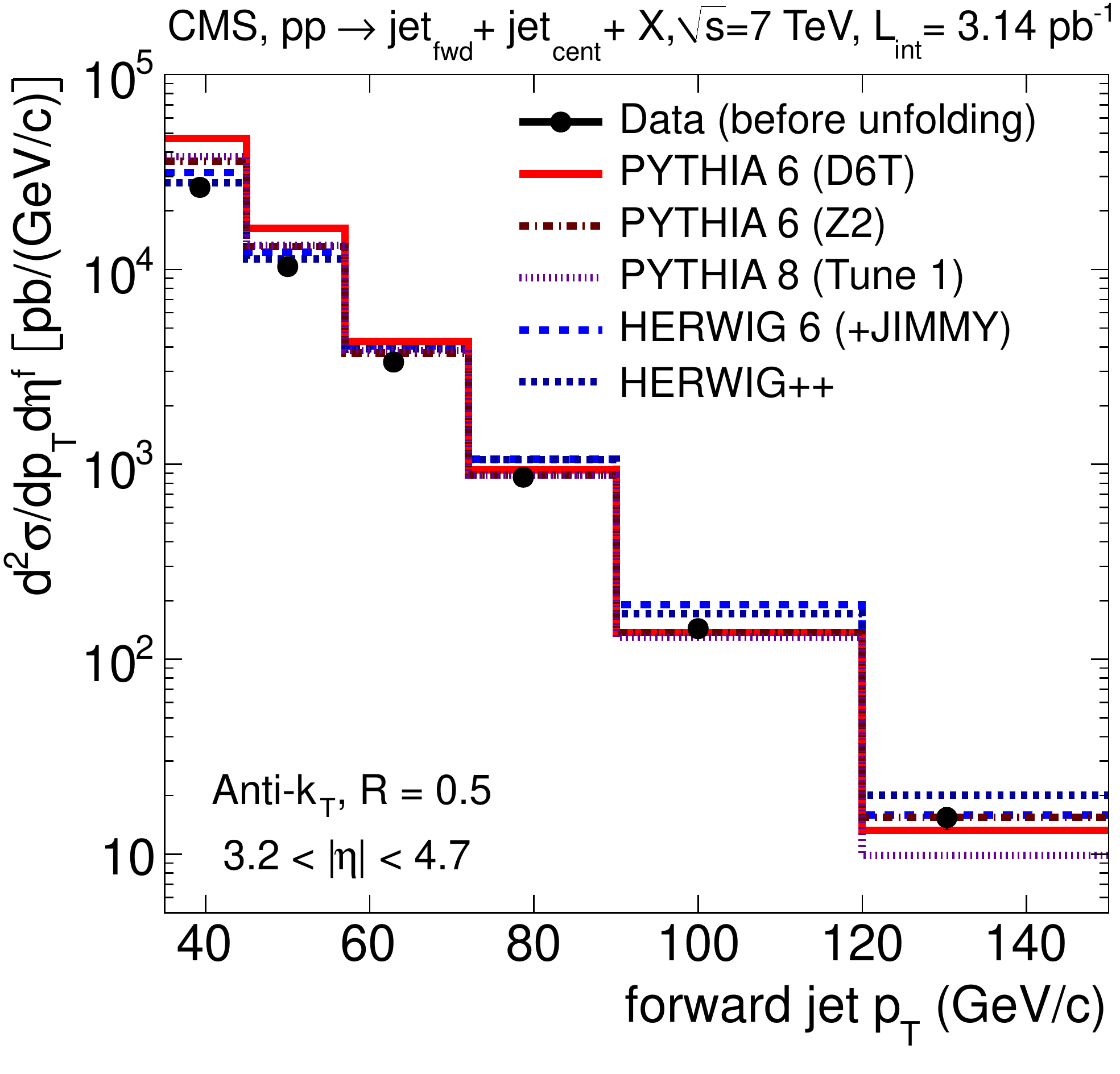}}
   \caption{Measured differential cross sections for jets as a function of $\pt$, before unfolding the energy resolution (black dots),
    compared to detector-level MC simulations generated with different versions of \PYTHIA\ and \HERWIG\ (histograms) for (a) inclusive forward
    jets, and for (b) central and (c) forward jets in dijet events.}
   \label{fig:jetsPtSpectrum}
\end{figure}

Figure~\ref{fig:jetsPtSpectrum} shows the reconstructed $\pt$ spectrum for: (a) inclusive forward jets,
and (b) central and (c) forward jets in dijet events. These are compared to MC events passed through full
detector simulation based on \GEANT~\cite{Agostinelli:2002hh}, and analysed in the same way as the data.
The data shown are calibrated through the JES normalisation, but not unfolded. 
The cross sections in each interval of $\eta$ and $\pt$ are divided by their bin-widths. 
With the simulated events normalised to the integrated luminosity used in this analysis,
all the models considered provide inclusive forward jet spectra consistent with the data,
but tend to overestimate the absolute cross sections measured in forward-central dijet events as discussed later.

The second correction (unfolding) of the measured jet spectrum is applied to account for the finite energy
resolution of the calorimeters. Although the bin size of the presented $\pt$ distributions equals or exceeds
the experimental resolution, the combination of a steeply falling $\pt$-spectrum and calorimeter resolutions
leads to migration of events across bins that can distort the distribution in $\pt$.
At central rapidities, the relative resolution in jet $\pt$, obtained from studies of $\pt$ imbalances in dijet events in data
and in MC simulations, changes from 15 to 8\% in the $\pt$ range of interest.
For forward jets, the relative resolution in $\pt$, derived from full-simulation studies and confirmed by the
momentum imbalance in dijet data, is below 12\% for $\pt >$~35\GeVc.
In the $\pt\approx$~35--80\GeVc range, the \textit{transverse} momentum resolution for HF jets is better than for central jets
because of the $\cosh(\eta)$-boost factor for the total energy deposited in the calorimeter at forward rapidities~\cite{Cerci:2008xv}.
Two methods are used to account for the bin-migration effect:
\begin{description}
\item (i) Exploiting the fact that MC simulations (Fig.~\ref{fig:jetsPtSpectrum}) reproduce the $\pt$-dependence of the inclusive forward-jet spectrum,
and that the simulated spectra for dijet events can be re-weighted to match the shape of the measured distributions, the MC samples
are used to study the bin-to-bin migrations. 
The correction factors have also been cross-checked by inverting the response matrix obtained from the MC
information, albeit with limited statistics, through the application of different unfolding algorithms~\cite{arXiv:1105.1160}.
\item (ii) The measured jet $\pt$~spectrum is fitted to some
parameterised function $f(\pt)$~\cite{ansatz-one,ansatz-two}, the result of which can be smeared using
the known (Gaussian) jet resolutions~\cite{Chatrchyan:2011me,Chatrchyan:2011ds}. The parameters of the model
are then defined by fitting the smeared transverse $\pt$ spectrum $F(\pt)$ to the measured $f(\pt)$, and
using the ratio of both distributions for the final correction~\cite{Chatrchyan:2011me}.
\end{description}

\begin{figure}[!Hhtb]
\subfigure[]{\includegraphics[width=0.333\textwidth]{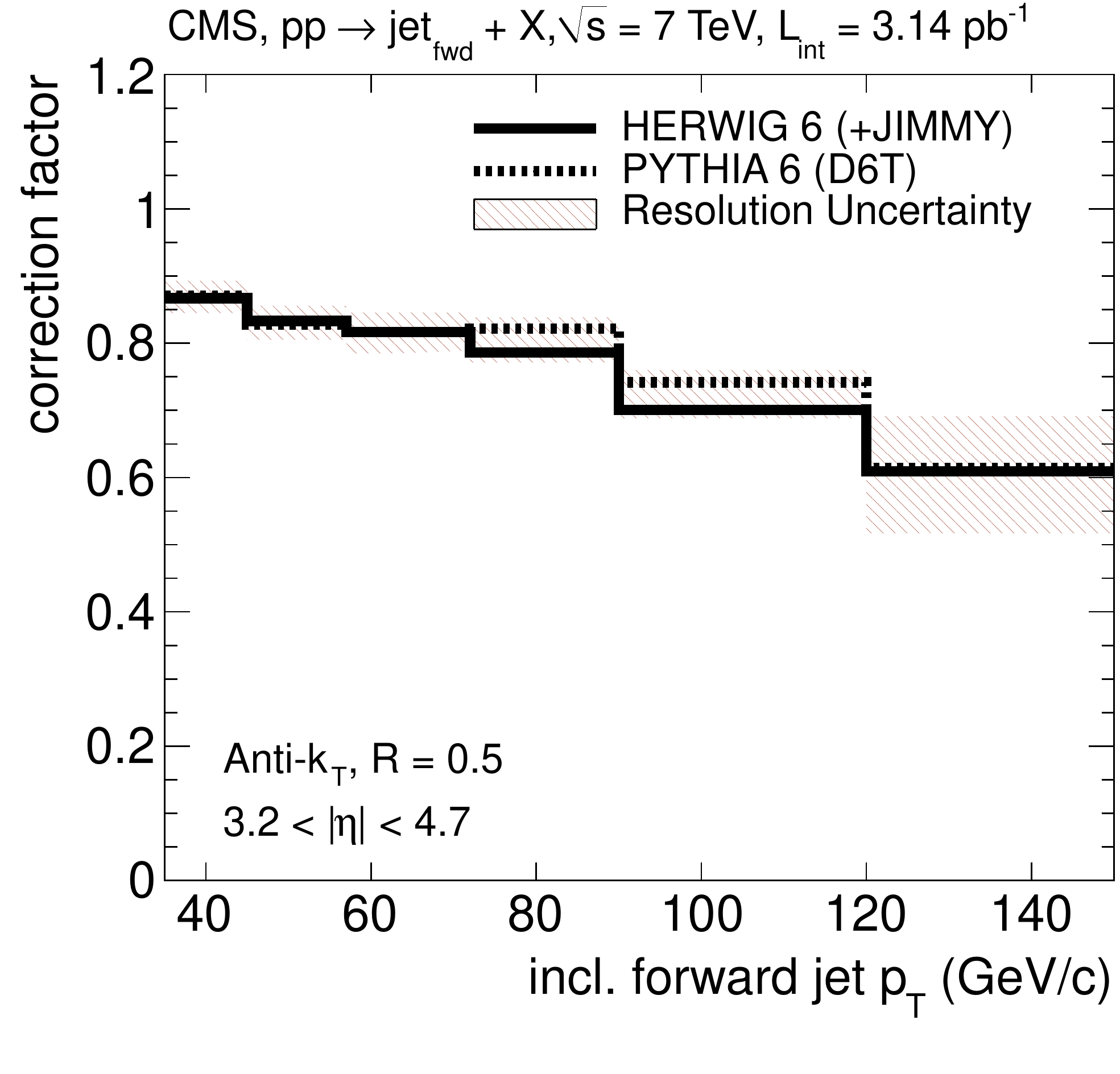}}
\subfigure[]{\includegraphics[width=0.333\textwidth]{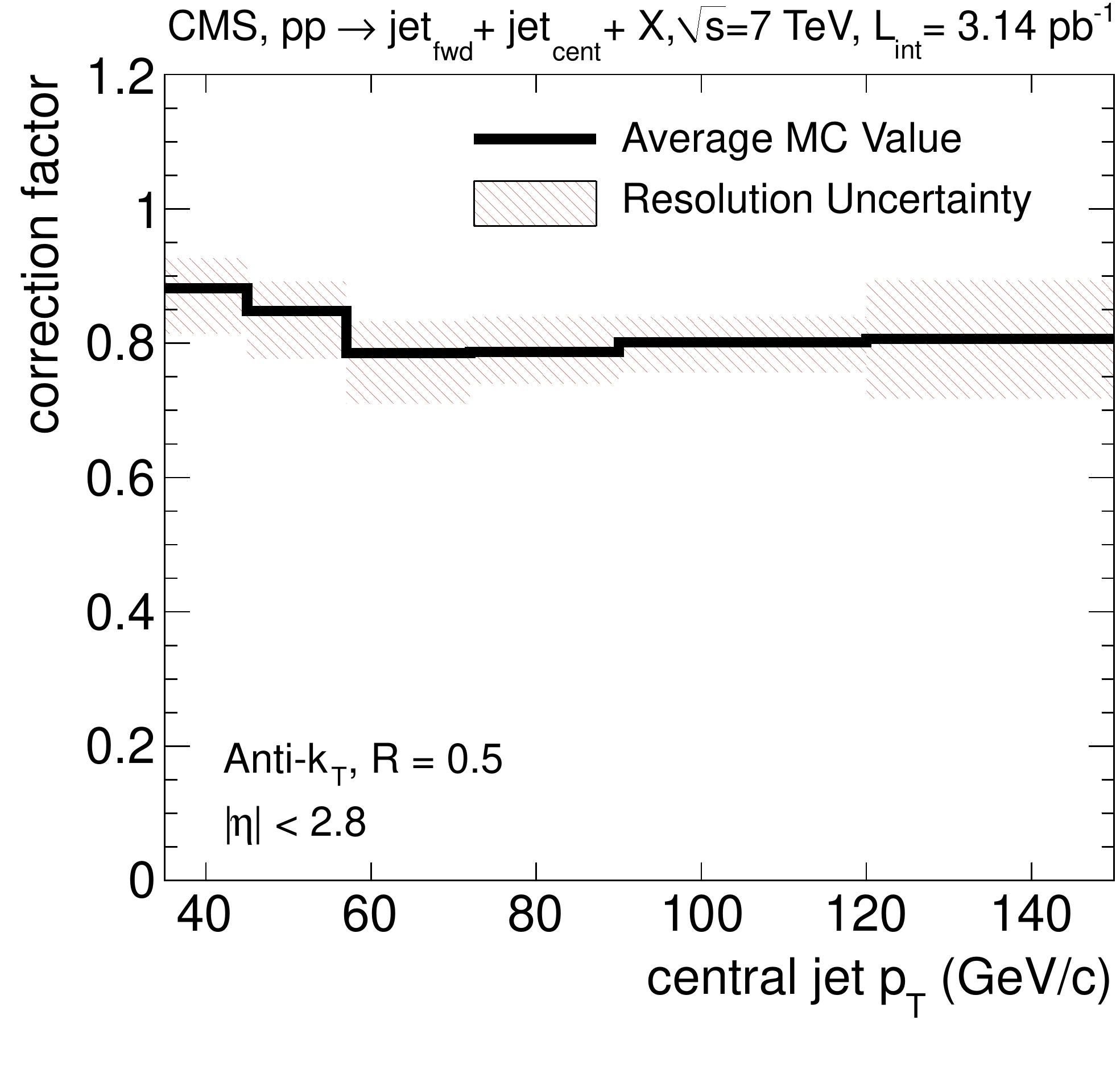}}
\subfigure[]{\includegraphics[width=0.333\textwidth]{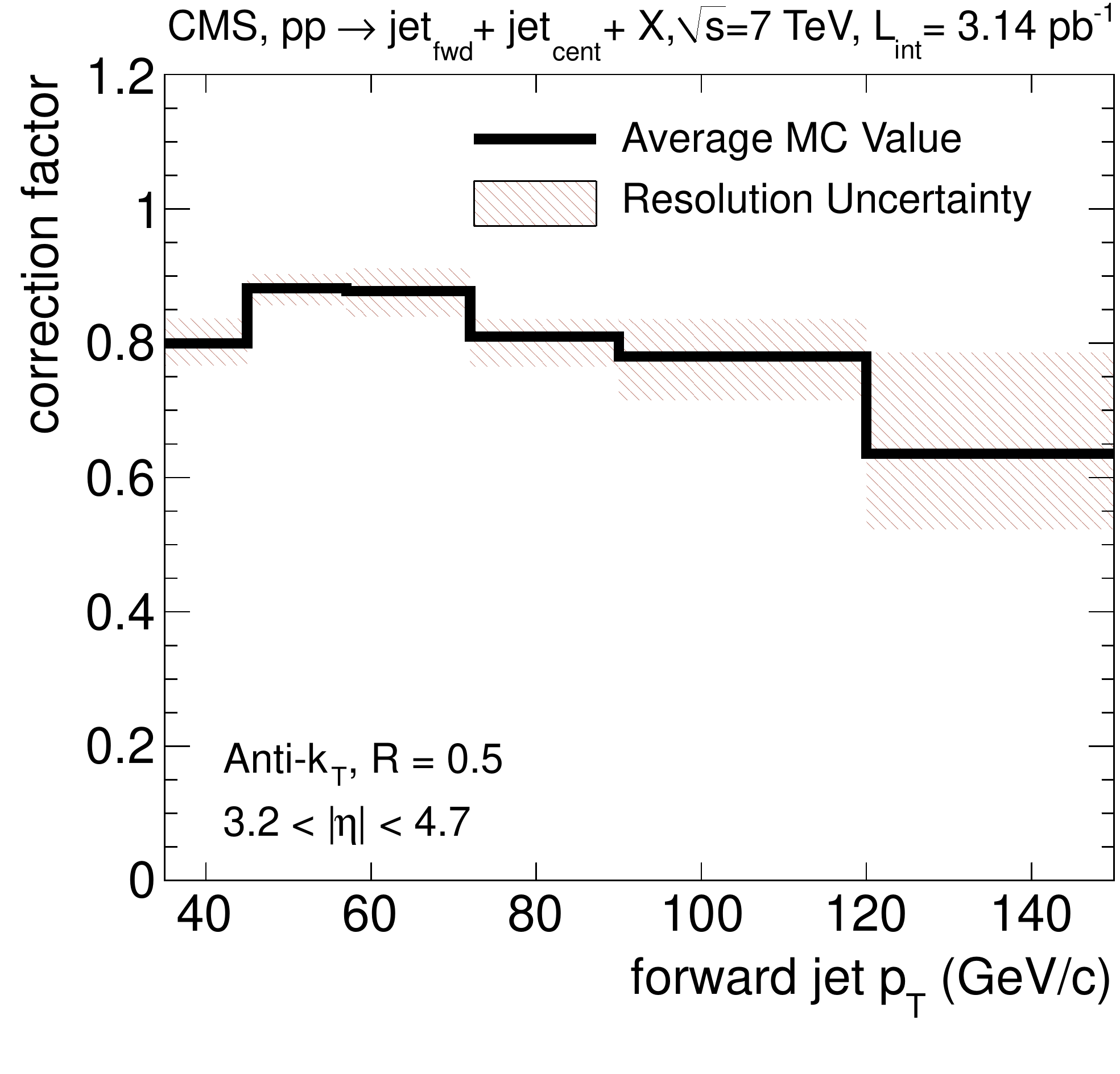}}
   \caption{The bin-by-bin unfolding correction factors as a function of $\pt$ for
    (a) inclusive forward jets, and for (b) central and (c) forward jets in dijet events.
    Panel (a) shows the individual correction factors obtained with \PYTHIA~6 and \HERWIG~6,
    while (b) and (c) show the average values obtained for the two MC generators (black histograms).
    The hatched band represents the uncertainties assigned to the correction factors as described in the text.}
   \label{fig:b2b_corr_factor}
\end{figure}

The difference between the results of the two methods is below 10\% for all $\pt$ bins.
The factors obtained with the MC method are used to correct the mean values of $\pt$,
while the results from the fits are used to assess the associated systematic uncertainties.
The MC-based method also takes into account various final-state effects,
such as hadronisation and particle decays, which affect the final energy clustered into jets.
The corresponding bin-by-bin factors thus fully correct the jet spectrum from the detector
to the particle levels via the factor
\begin{equation}
C_{\text{had}}(\pt,\eta) = \frac{N^{\mathrm{MC}}_{\text{had}}(\pt,\eta)}{N^{\mathrm{MC}}_{\text{det}}(\pt,\eta)}~,
\label{eq:correction}
\end{equation}
where $N^{\mathrm{MC}}_{\text{had}}(\pt,\eta)$ and $N^{\mathrm{MC}}_{\text{det}}(\pt,\eta)$ are the jet
event yields determined after hadronisation and after full simulation, respectively.
The factor $N^{\mathrm{MC}}_{\text{had}}(\pt,\eta)$ is obtained by
averaging the predictions from \PYTHIA~6 with \HERWIG~6+\textsc{Jimmy}, which provide different modelling of
parton-to-hadron processes, one based on string and the other on cluster fragmentation, respectively.
The unfolding correction factors obtained for the two event generators differ by less than 5\%, as shown in the left panel of
Fig.~\ref{fig:b2b_corr_factor}. (The average of the two MC predictions is shown in the two right panels.)
The hatched band in all panels indicates the uncertainty obtained by changing the jet $\pt$
resolution by $\pm$10\%, and covers the range of differences found for the two methods of unfolding the data.

\section{Determination of jet cross sections and systematic uncertainties}
\label{sec:spectra}

The final data sample contains events with at least one forward jet or at least a forward and a central
jet satisfying the selections described in Section~\ref{sec:sample_evt_select}.
The corresponding numbers of events, $N_\text{evts}$, are binned into
a differential inclusive jet cross section as a function of $\eta$ and $\pt$:
\begin{eqnarray}
\frac{\rd^2\sigma}{\rd\pt\,\rd\eta} = \frac{ C_{\text{had}}}{\mathcal{L} \cdot \varepsilon_t}\cdot\frac{N_\text{evts}}{\Delta \pt \cdot\Delta \eta}\; .
\label{eqn:x-sect-formula}
\end{eqnarray}
The factor $C_{\mathrm{had}}$ accounts for bin-to-bin migrations due to $\pt$ resolution
and detector to particle corrections, Eq.~(\ref{eq:correction}), ${\cal L}$ is the total integrated
luminosity, $\varepsilon_t$ is the efficiency of the jet trigger, and
$\Delta \pt $ and $\Delta \eta$ are the sizes of the bins in $\pt $ and $\eta$, respectively.
The jet triggers have a $\varepsilon_t$~=~100\% efficiency for all $\pt$ and $\eta$ values
considered in this paper 
with a negligible contribution to the total systematic uncertainty.

There are three primary sources of systematic uncertainty in the jet cross sections measurements:
\begin{description}
\item (i) Jet energy scale in the calorimeters. At forward rapidities, the HF
calorimeter has a JES calibration uncertainty that varies between 3 and 6\%, depending on the $\pt$ and $\eta$
of the reconstructed jet~\cite{Chatrchyan:2011ds}. This uncertainty must be convoluted with that associated
with a $\sim$0.8\GeV energy shift per pileup-event due to the presence of other hadrons around the forward jet axis.
The JES uncertainties of the central calorimeters have typical values between 2.5 and 3.5\%~\cite{Chatrchyan:2011ds}.
The uncertainty from pileup energy has been studied by comparing central jet $\pt$ distributions
with and without the requirement to have only one primary vertex in the event.
The central jet $\pt$ spectra under these two conditions are found to
differ by less than 5\%. The JES uncertainties, propagated to the steeply falling jet spectra
(inverse power-law $\pt$ distributions with exponent of $n\approx$~5), translate into
uncertainties of the order of $\pm($20--30)\% in the final forward and central jet cross sections.
\item (ii) Unfolding procedure and $\pt$ resolution ($C_{\mathrm{had}}$ factor).
The $\pm 10\%$ uncertainty on the jet $\pt$ resolution (Fig.~\ref{fig:b2b_corr_factor})
translates into an uncertainty of 3 to 6\% (increasing with $\pt$) on the final cross sections.
An additional uncertainty of 3\%, from the model dependence of the correction factors defined by
the difference between the \PYTHIA~6 and \HERWIG~6 generators used to unfold the
cross sections, is added in quadrature.
\item (iii) Luminosity. The uncertainty of the integrated pp luminosity
results in a 4\% uncertainty on the overall normalisation of the spectra~\cite{CMS-DP-2011-002}.
\end{description}

\ifx\ver\verALL
 \begin{figure}[!htb]
  \centering
  \subfigure[]{\includegraphics[width=0.333\linewidth]{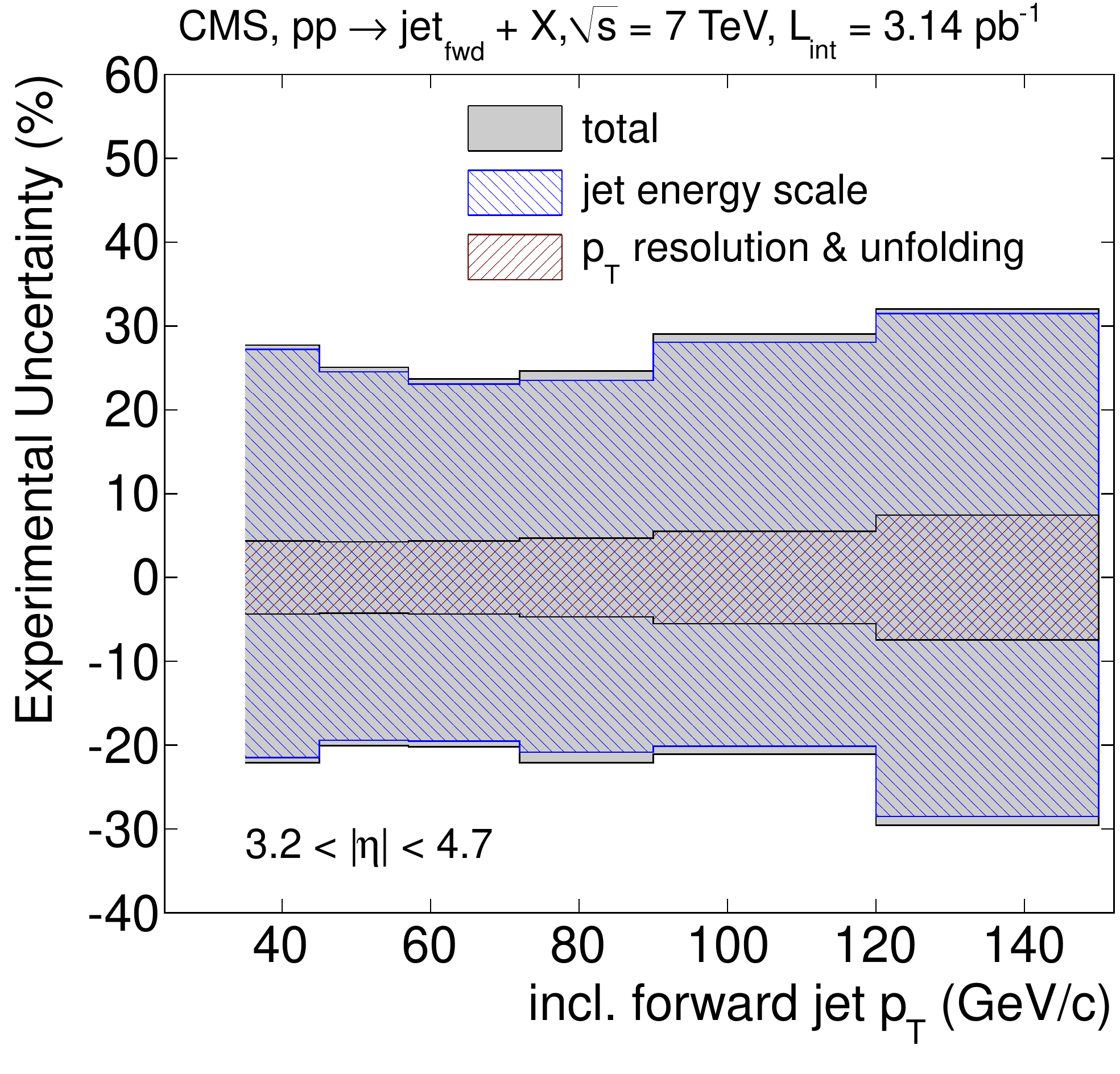}}
  \subfigure[]{\includegraphics[width=0.333\textwidth]{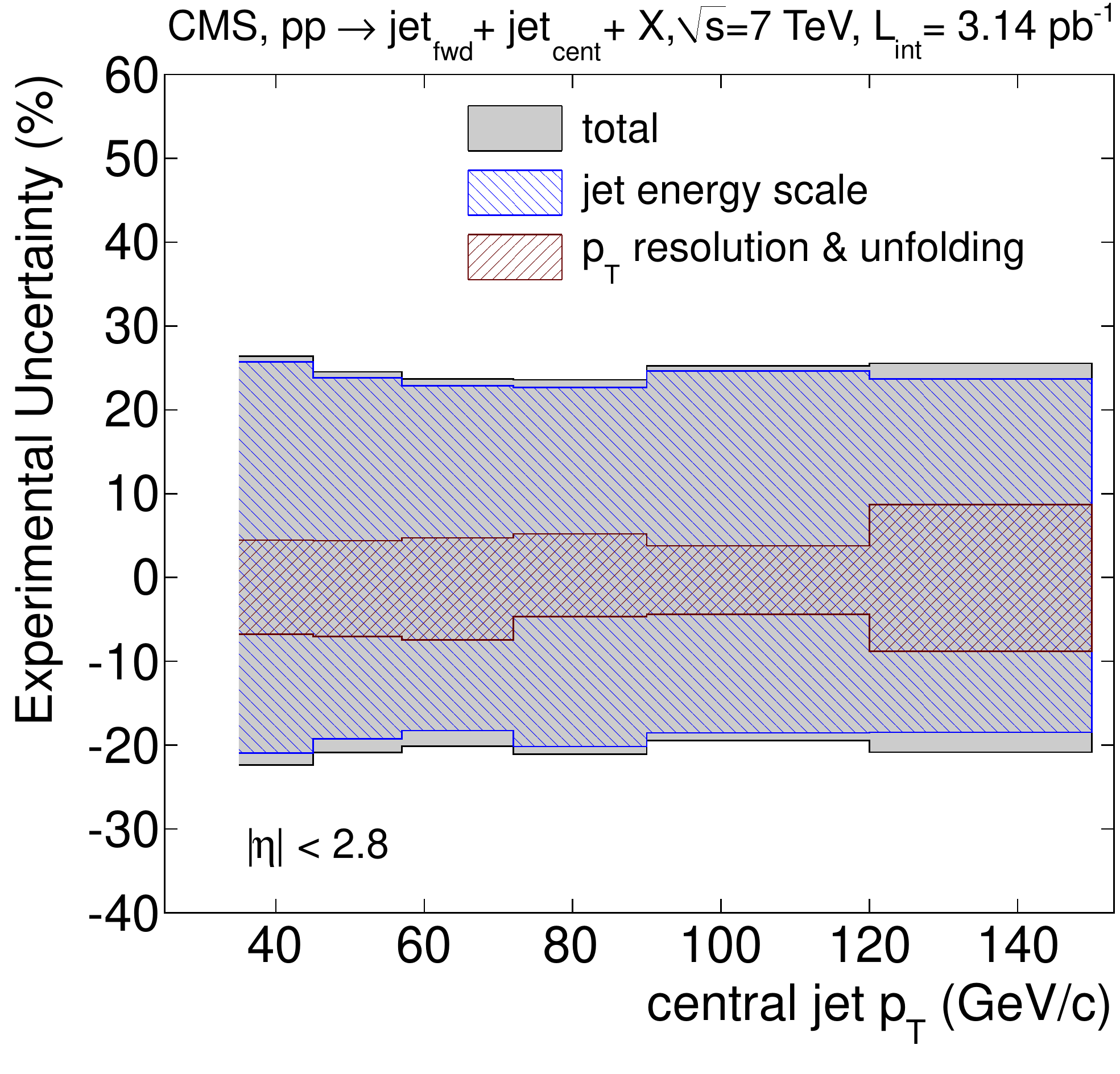}}
  \subfigure[]{\includegraphics[width=0.333\textwidth]{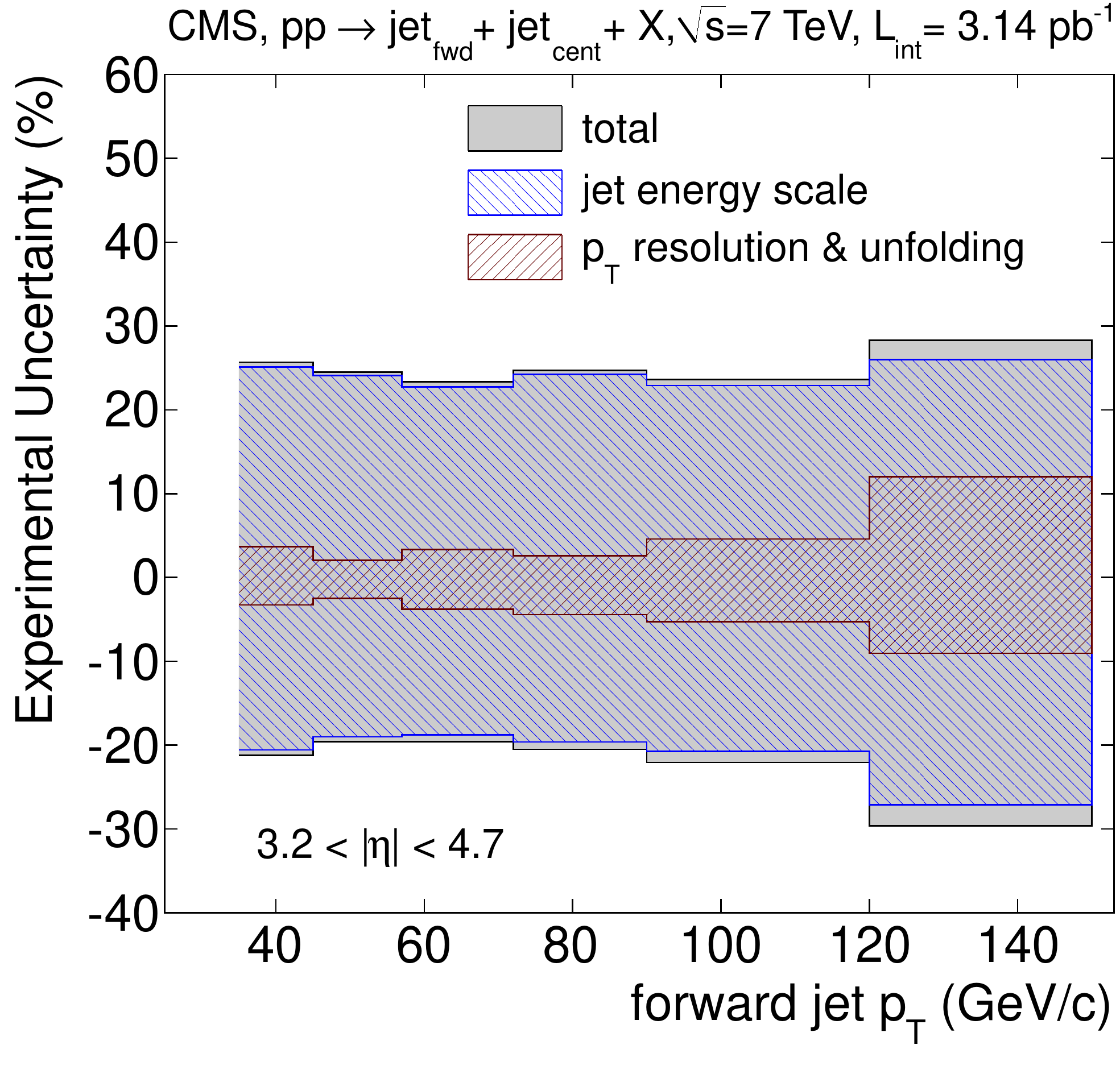}}
    \caption{Systematic uncertainties as a function of jet $\pt$ for (a) inclusive forward production,
     and for (b) central and (c) forward jet spectra in dijet events.
     The outer limits of the grey areas show the overall uncertainties, from adding in quadrature uncertainties
     from the JES, the unfolding and the luminosity.}
   \label{fig:ExpUncertainty}
 \end{figure}
\fi

In all $\pt$ bins of the measured cross sections, the statistical uncertainty (of the order of 1--2\% in the
low $\pt$ bin and 5--10\% in the highest) is small relative to the systematic uncertainty obtained by adding
all contributions in quadrature. The latter amounts to $\sim$30\% and is dominated by the uncertainty on the
calibration of the jet energy scale. The different contributions to the systematic uncertainty are shown as a
function of jet $\pt$ in Fig.~\ref{fig:ExpUncertainty} for the three $\pt$ distributions of interest.
The grey areas show the total uncertainty, while the two hatched areas indicate the uncertainties on the JES
and the unfolding procedure. Table~\ref{tab:sigmaexp} tabulates the measured, fully-corrected
$\pt$-differential jet cross sections and their associated uncertainties.

\begin{table}[!htbp]
\caption{Measured $\pt$-dependent differential cross sections for inclusive forward jets (second column), and for central (third column)
  and forward (last column) jets in dijet events. The first (second) uncertainty reflects the statistical (systematic)
 contribution.
\label{tab:sigmaexp}}
\begin{center}\setlength{\extrarowheight}{1pt}
\begin{tabular}{cccc}\hline
$\pt$ bin (centre) &  $\frac{\rd^2\sigma}{\rd\pt\,\rd\eta}$  & $\frac{\rd^2\sigma}{\rd \pt^c\,\rd \eta^c}$ & $\frac{\rd^2\sigma}{\rd\pt^f\,\rd\eta^f}$\\
\GeVc &  pb/(\GeVcns) & pb/(\GeVcns) & pb/(\GeVcns) \\\hline \\
35 $-$ 45 (39.3) & $\left(89 \pm 0.2^{+24}_{-19}\right)\times 10^{3}$ & $\left(10 \pm 0.1^{+2.6}_{-2.2}\right)\times 10^3$& $\left(21 \pm 0.2^{+5.4}_{-0.5}\right)\times 10^3$ \\
45 $-$ 57 (50.2) & $\left(20 \pm 0.1^{+4.9}_{-3.9}\right)\times 10^{3}$ & $\left(5.2 \pm
  0.07^{+1.2}_{-1.0}\right)\times10^3$ & $\left(9.2 \pm 0.1^{+2.2}_{-1.8}\right)\times 10^3$\\
57 $-$ 72 (63.2) & $\left(4.4 \pm 0.04^{+1.0}_{-0.9}\right)\times 10^{3}$ & $\left(1.9 \pm
  0.03^{+0.4}_{-0.4}\right)\times10^3$ & $\left(2.9 \pm 0.06^{+0.7}_{-0.6}\right)\times 10^3$\\
72 $-$ 90 (79.2) & $880 \pm 10^{+200}_{-180}$ & $590 \pm 20^{+130}_{-120}$ & $690 \pm 30^{+170}_{-140}$ \\
90 $-$ 120 (101.0) & $115 \pm 4^{+40}_{-25}$ & $135 \pm 6^{+33}_{-25}$ & $110 \pm 8^{+25}_{-25}$\\
120 $-$ 150 (132.0) & $10 \pm 1.2^{+3}_{-3}$ & $28 \pm 3^{+7}_{-5}$ & $10 \pm 2.3^{+3}_{-3}$\\ \\ \hline
\end{tabular}
\end{center}
\end{table}

\section{Results and comparison to theory}
\label{sec:data_vs_theory}

\subsection{Theoretical predictions}

The measured differential jet cross sections are compared to predictions from different pQCD approaches:
(i) general-purpose event generators \PYTHIA~6 (version 6.422) with D6T and Z2 tunes~\cite{pythia,tune},
\PYTHIA~8 (version 8.135) with Tune~1~\cite{Sjostrand:2007gs}, \HERWIG~6 (version 6.510.3)~\cite{herwig6}
with underlying-event modelled with \textsc{Jimmy}~\cite{jimmy}, and \textsc{herwig++} (version 2.3)~\cite{Bahr:2008tx},
(ii) NLO calculations obtained with the {\sc powheg} package~\cite{Alioli:2010xa} (matched with
\PYTHIA\ and \HERWIG\ parton showers) as well as with \textsc{nlojet++}~\cite{Nagy:2001fj} within the
\textsc{fastNLO}~\cite{Kluge:2006xs} package, for different sets of parton densities,
and (iii) the \textsc{cascade} (version 2.2.04)~\cite{Deak:2009xt,Deak:2010gk} and \textsc{hej} \cite{Andersen:2009nu,Andersen:2011hs} codes.

The \PYTHIA\ and \HERWIG\ Monte Carlo event generators are based on standard collinear (DGLAP)
evolution, where the parton shower can be developed by ordering the parton splittings in $\pt$
(in the Z2 tune, in Tune 1, or in combination with the \POWHEG\ NLO generator) or in virtuality $Q^2$
(in the D6T tune). \HERWIG\ uses angular ordering for shower evolution.
The \PYTHIA~6 and \HERWIG~6 event generators use the CTEQ6L PDF~\cite{Pumplin:2002vw}, whereas
CTEQ5L~\cite{Lai:1999wy} has been used for \PYTHIA~8, and the MRST2001 PDF~\cite{Martin:2001es} for \textsc{herwig++}.
The default \textsc{nlojet++} calculation is run with CT10~\cite{ct10}, and \POWHEG\ is run with the
CTEQ6M PDF~\cite{Pumplin:2002vw} plus \PYTHIA~6 (Perugia 0 tune~\cite{P0}) and \HERWIG~6 for the
parton showering and hadronisation. The default renormalisation and factorisation scales
have been set to $\mu_\mathrm{r} = \mu_\mathrm{f} = \pt$ for both NLO calculations.
The \textsc{cascade} Monte Carlo program, based on resummation of leading logarithms in virtuality $Q^2$ and in parton momentum fraction $x$, as implemented in the
CCFM evolution equations, uses the Set-A unintegrated parton distributions~\cite{Jung:2004gs}
and a cut on  the $\pt$ of the matrix-element partons of 14\GeVc.
The \textsc{hej} event generator uses the MSTW2008NLO PDFs~\cite{Martin:2009iq} and provides, at parton level, an
all-order description of the dominant radiative corrections for hard,
wide-angle emissions. 

\begin{figure}[!Hhtb]
\begin{center}
 \subfigure[]{\includegraphics[width=0.499\textwidth,clip=true]{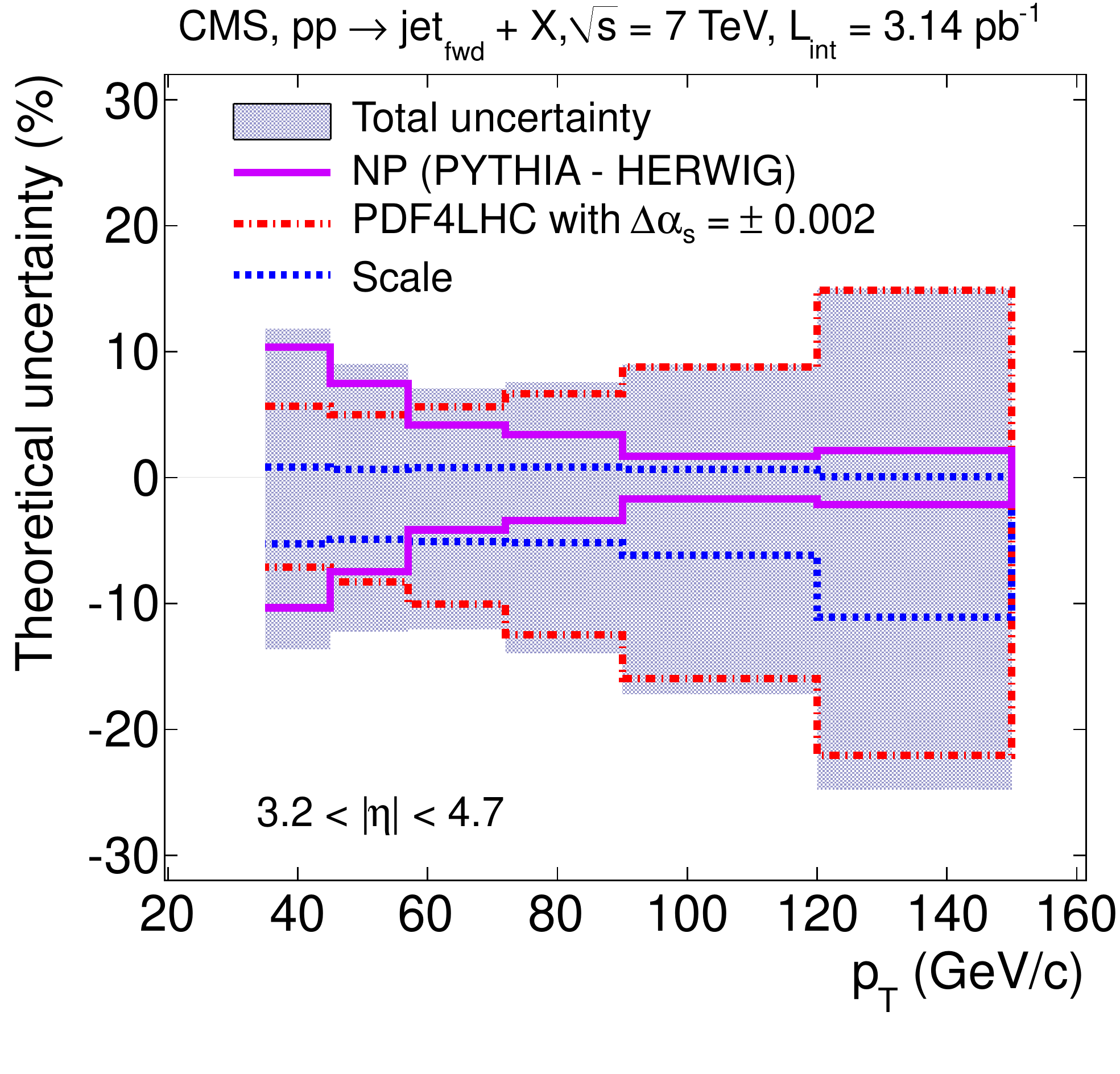}}
 \subfigure[]{\includegraphics[width=0.499\textwidth,clip=true]{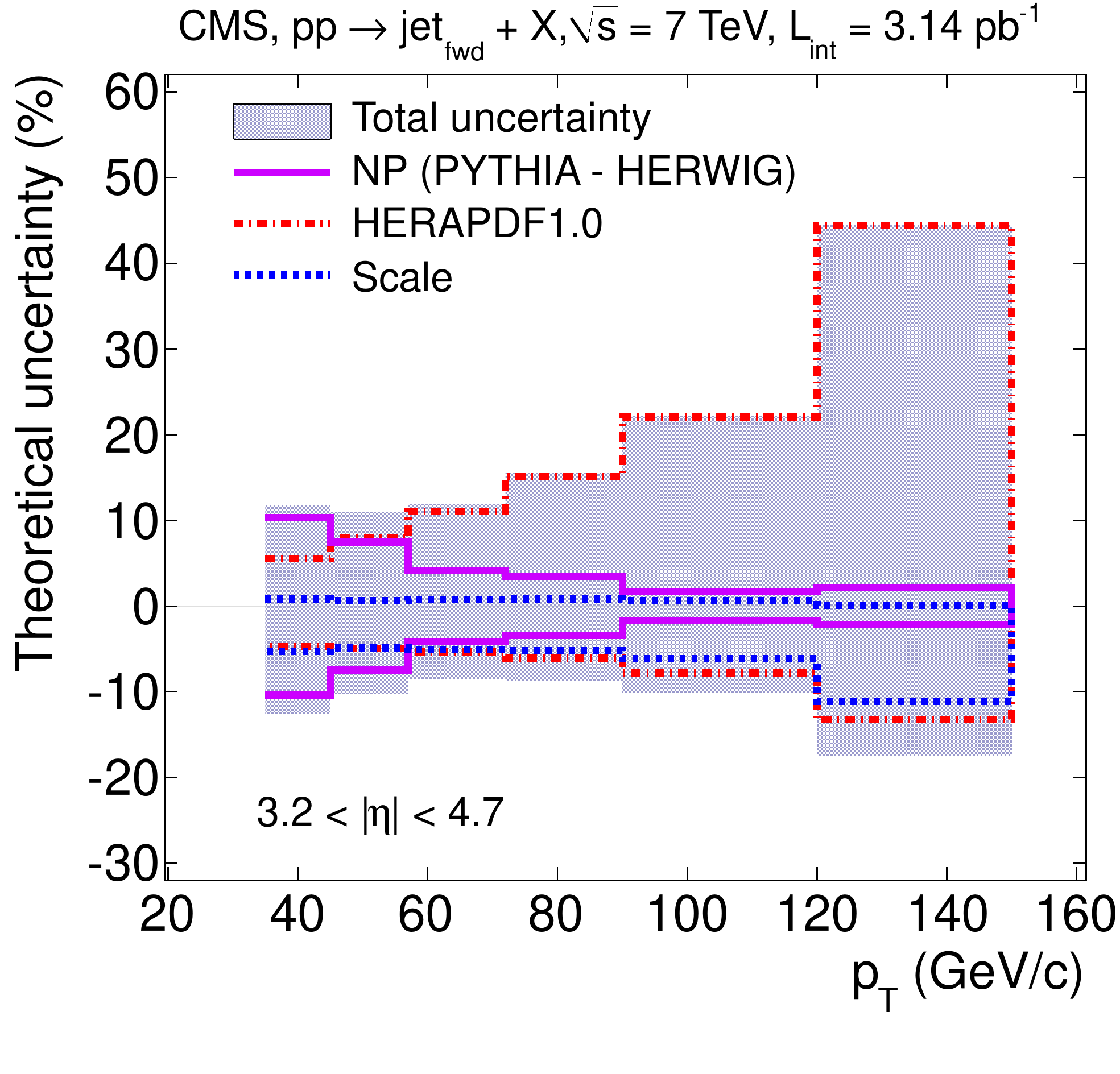}}
\end{center}
  \caption{
    Uncertainties on the predicted NLO inclusive forward jet spectrum. Plot (a) shows the contributions from
    non-perturbative effects, choice of
    PDF and the value of the strong coupling $\alpha_{S}$ (computed with the PDF4LHC prescription),
    and uncertainties associated with the renormalization and fragmentation scales. Plot (b) shows the uncertainties from NP,
    PDF and $\alpha_{S}$ (obtained with HERAPDF1.0), and the theoretical scales. Total uncertainties are obtained
    by adding quadratically the uncertainties on NP, PDF and the scales.}
  \label{fig:th_uncert}
\end{figure}

Before comparing the data to parton-level predictions such as \textsc{nlojet++} or \textsc{hej},
the uncertainties from non-perturbative (NP) effects must be determined.
Such effects include energy lost from the jet in the hadronisation process or ``splashed-in''
from the underlying event (UE) into the jet,
and are estimated as in Ref.~\cite{Abazov:2008hu}, by comparing the \PYTHIA~6 and \HERWIG~6+\textsc{Jimmy}
parton-level spectra with the corresponding particle-level predictions after hadronisation and UE activity.
Each MC program has a different way of modelling parton hadronisation and multiparton interactions, that
results in different UE characteristics. The NP correction factors amount to 1.10 (1.02) at the lowest
(highest) $\pt$ bin considered in this study.
Half of the difference between these two predictions, displayed as a function of forward jet $\pt$ in
Fig.~\ref{fig:th_uncert}, is taken as an estimate of the total systematic uncertainty associated with this NP effect.

For NLO predictions (\textsc{nlojet++} and \POWHEG), the uncertainties associated with the
PDF and the strong coupling $\alpha_{S}$ can be estimated following the PDF4LHC interim recommendation~\cite{pdf4lhc}.
The uncertainty on the PDF is estimated from the maximum envelope obtained from the 68\% confidence-level
eigenvectors (CL68) of the CT10, MSTW2008 and NNPDF2.1~\cite{Ball:2011mu} sets. The uncertainty from the value of the strong
coupling $\alpha_{S}$ is derived from separate fits using the CT10 PDF, where $\alpha_{S}(M_{Z})$ is
changed by $\pm 0.002$, and is added in quadrature to the uncertainty on the PDF. The uncertainty
associated with higher-order corrections neglected in the NLO calculation has been evaluated by changing
the renormalisation and factorisation scales by factors proportional to the jet $\pt$
in the following six combinations: 
($\mu_{\rm r},\mu_{\rm f}$)~=~($\pt$/2,$\pt$/2), ($\pt$/2,$\pt$), ($\pt$,$\pt$/2), ($\pt$,2$\pt$), (2$\pt$,$\pt$) and (2$\pt$,2$\pt$)~\cite{point6}.
Figure~\ref{fig:th_uncert} (a) shows all the sources of theoretical uncertainty.
The NP corrections dominate for $\pt <$~60\GeVc, whereas uncertainties on PDF and $\alpha_{S}$ dominate
above that $\pt$. Scale uncertainties are less important at all transverse momenta.
These three sources of uncertainty are added in quadrature into a single band representing the
NLO theoretical uncertainty.

An independent cross-check of the uncertainty due to the PDF choice is given in Fig.~\ref{fig:th_uncert}(b), which shows the
same uncertainties for NP and scale, but with the PDF envelope obtained by using the HERAPDF1.0 parton
densities~\cite{HERAPDF10:2009wt}. The 33 HERAPDF1.0 PDF eigenvalues correspond to 68\% CL
intervals of this PDF that account for experimental, model and parametrisation uncertainties on
the fit to HERA data. Two more HERAPDF1.0 fits, with $\alpha_s$ changed by $\pm 1$ standard deviation of the
world-average value (0.1176$\pm$0.002)~\cite{Amsler:2008zzb}, are also checked, and the corresponding effect added in
quadrature to the PDF uncertainty. For jets at high $\pt$, this uncertainty is larger than the one
obtained with the PDF4LHC prescription, as the HERAPDF1.0 sets have fewer constraints on the gluon density at
high-$x$ than other globally-fitted PDF, and because HERAPDF also includes extra
uncertainties on the initial shape of the parton distributions.

\subsection{Inclusive forward spectrum}

The fully corrected inclusive forward jet cross section as a function of $\pt$ is shown in
Fig.~\ref{fig:xsection}(a) compared to the models discussed above. The data points are plotted at the ``true'' centre of the $\pt$ distribution in that bin~\cite{Lafferty:1994cj}.
The experimental systematic uncertainty (Fig.~\ref{fig:ExpUncertainty}) is
shown as a grey band. Figure~\ref{fig:xsection}(b) shows the ratio of theoretical to
experimental jet cross sections, including the NLO band of uncertainty
(Fig.~\ref{fig:th_uncert}). Within the theoretical and experimental uncertainties,
the predictions are in good agreement with the measurements.

\ifx\ver\verALL
\begin{figure}[!Hhtb]
 \begin{center}
  \subfigure[]{\rotatebox{0}{\includegraphics[width=0.49\textwidth]{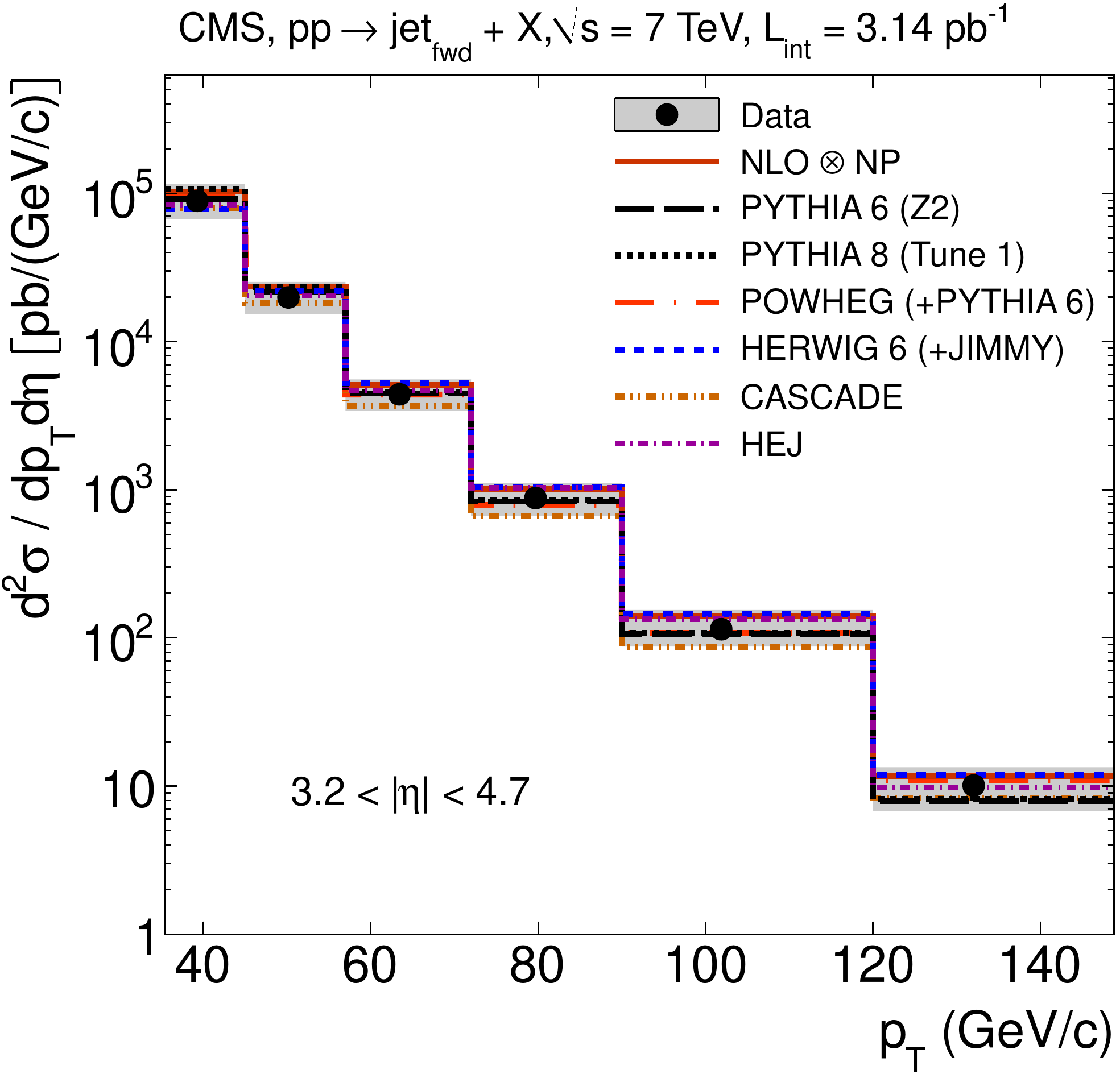}}}\vspace{0.1cm}
  \subfigure[]{\rotatebox{0}{\includegraphics[width=0.49\textwidth]{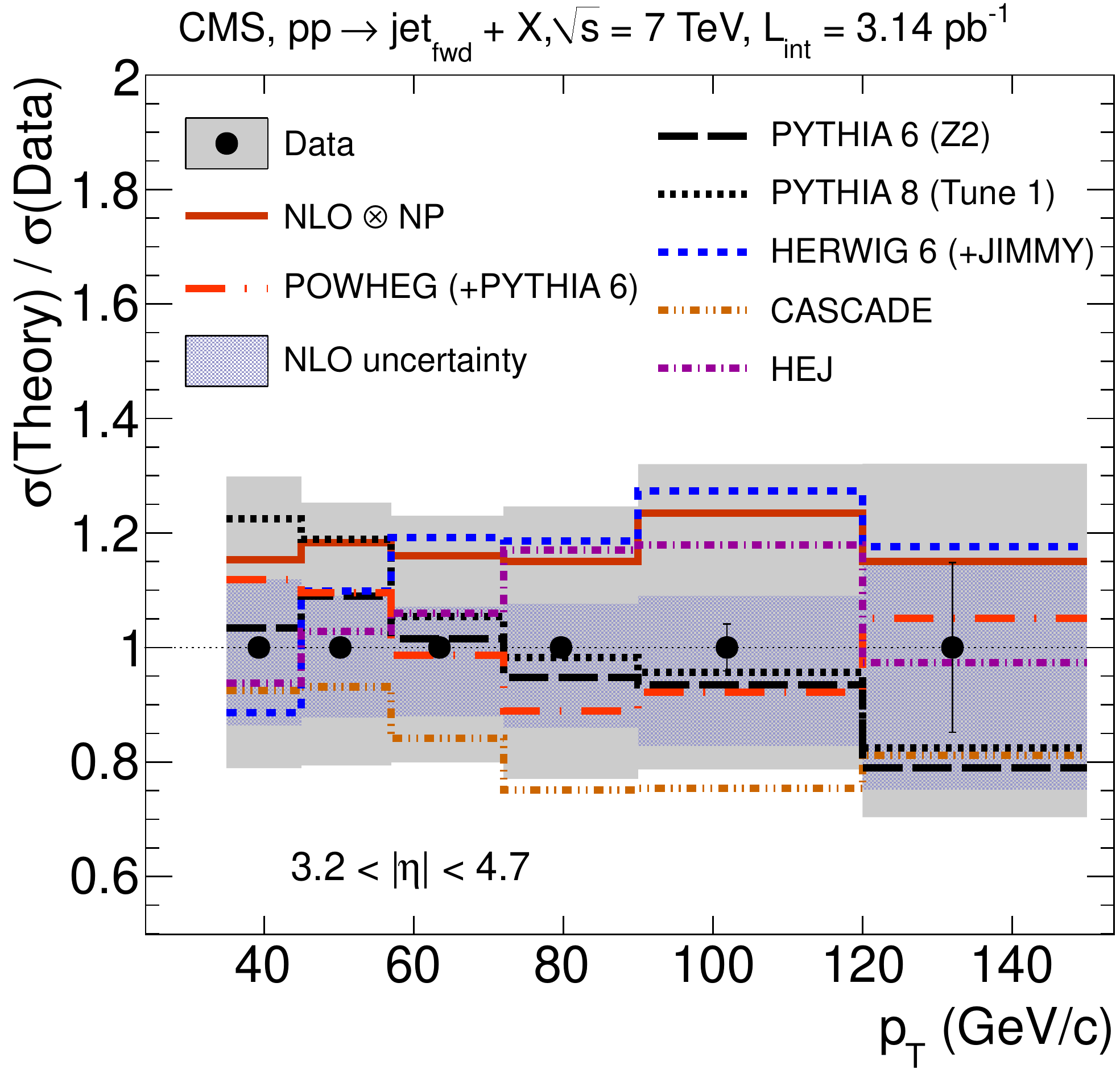}}}
  \caption{(a) Inclusive jet cross section at forward pseudorapidities
    (3.2~$<|\eta|<$~4.7), fully corrected and unfolded, compared to
    particle-level predictions from \PYTHIA~6, \PYTHIA~8, \HERWIG~6, \textsc{nlojet++} corrected for
    non-perturbative effects, \POWHEG, \textsc{cascade} and \textsc{hej}.
    (b) Ratio of theory/data for the forward jet spectrum.
    The error bars on all data points (which, in (a), are smaller than the size of the markers)
    reflect just statistical uncertainties, with systematic uncertainties plotted as grey bands.
    The dark band in (b) shows the theoretical uncertainty on the NLO predictions.}

  \label{fig:xsection}
 \end{center}
\end{figure}
\fi

\begin{figure}[!htbp]
   \centering
   \subfigure[]{\includegraphics[width=0.333\textwidth,clip=true]{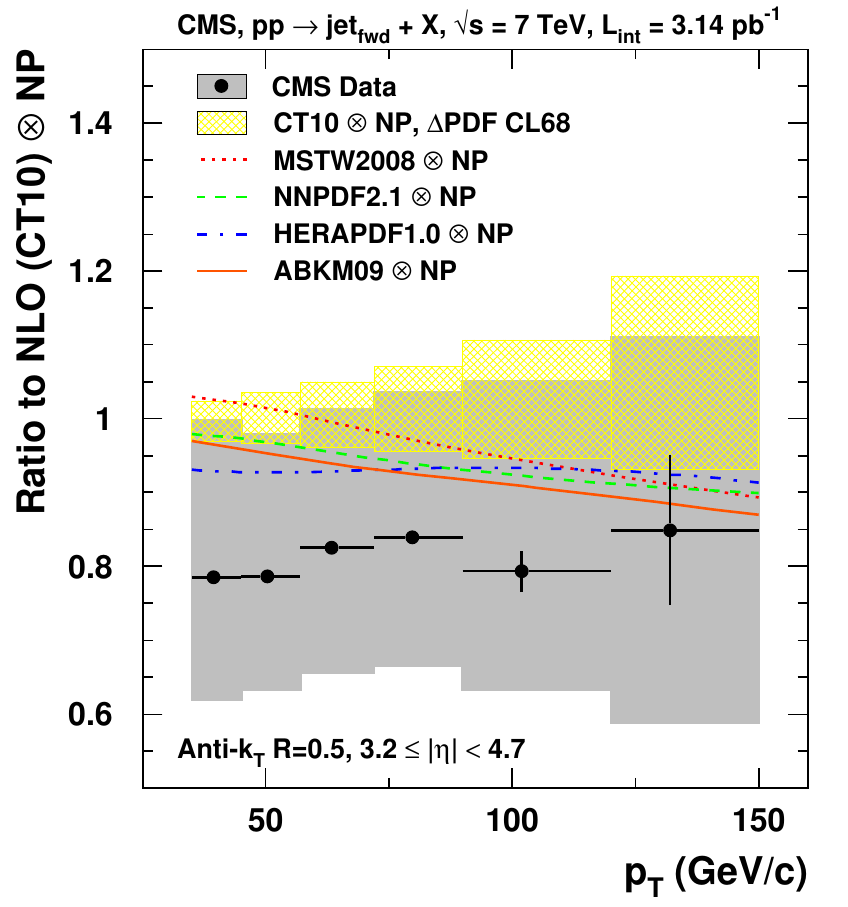}}
   \subfigure[]{\includegraphics[width=0.333\textwidth,clip=true]{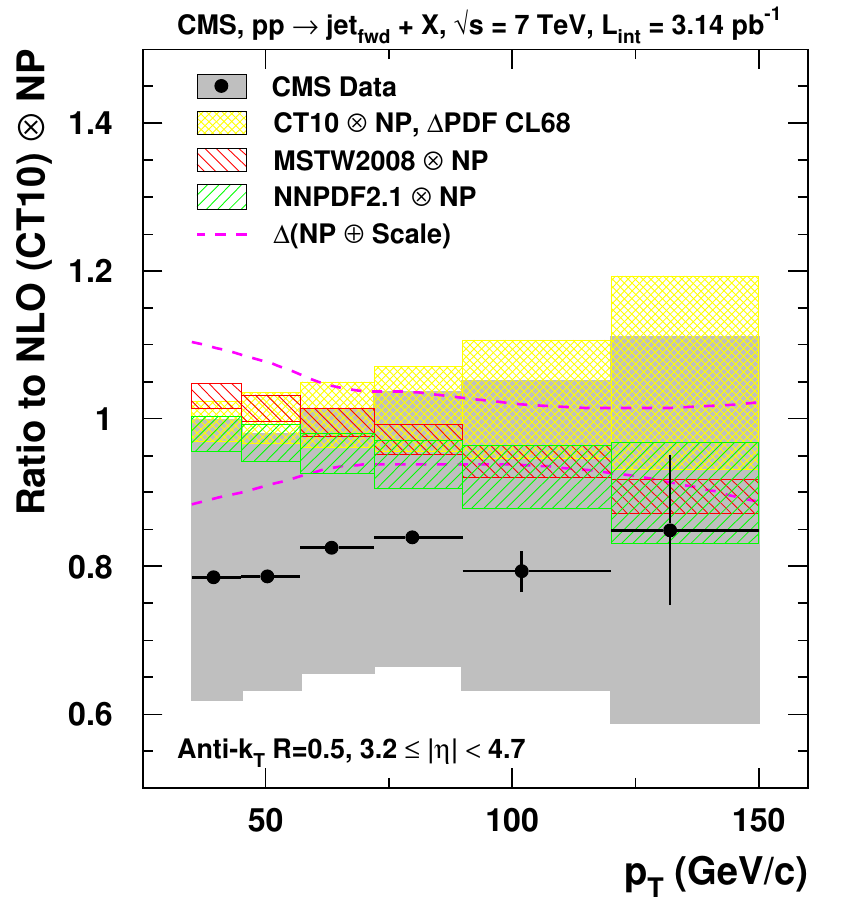}}
   \subfigure[]{\includegraphics[width=0.333\textwidth,clip=true]{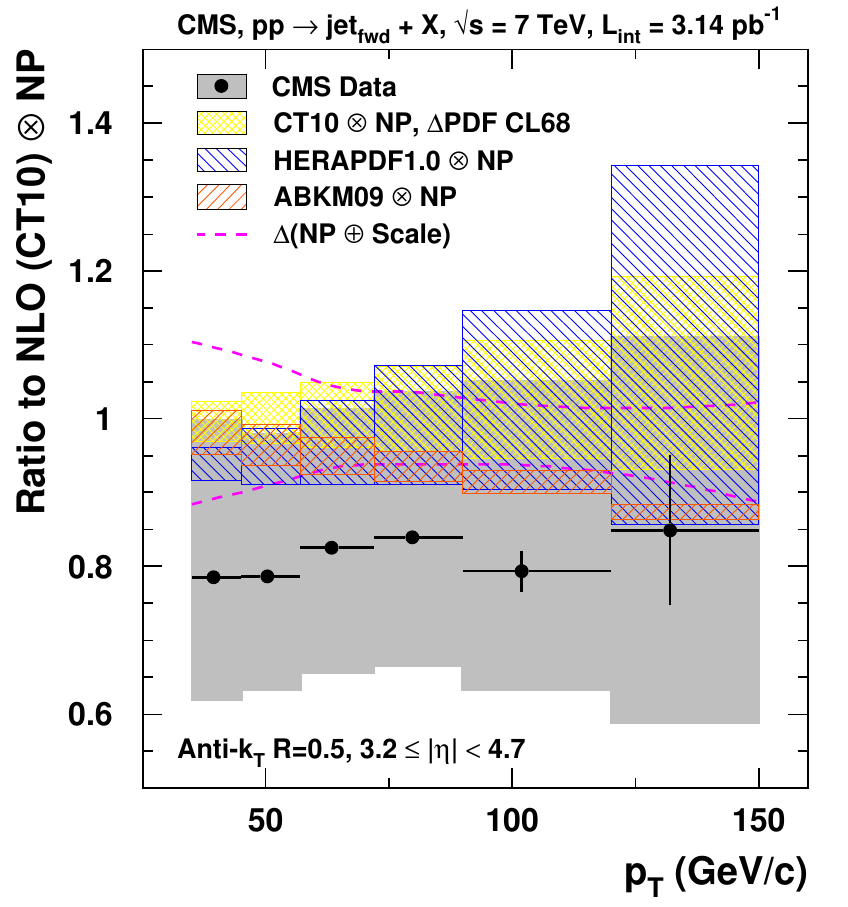}}
   \caption{Ratio of the inclusive forward jet spectrum for data over the NLO predictions using the CT10 PDF,
    as a function of $\pt$, shown with the statistical (error bars) as well as systematic uncertainties (grey band).
    Additional predictions are shown in (a) for all the central PDF predictions (curves),
    (b) for the MSTW2008 and NNPDF2.1 sets, and in (c) for the HERAPDF1.0 and ABKM09 PDF.
    The corresponding PDF uncertainties are shown as coloured bands around the ratios.
    Common theoretical uncertainties from choices of scale and non-perturbative corrections
    are indicated by dashed (magenta) curves in (b) and (c).}
   \label{fig:fwdjets_vs_PDFs}
\end{figure}

To gauge the sensitivity of the forward jet measurement to the underlying parton densities in the
proton, Fig.~\ref{fig:fwdjets_vs_PDFs} shows the NLO predictions compared to the data in the form of
bin-by-bin ratios of data to theory (which is used instead of theory/data to improve graphical presentation
at high $\pt$ where the reference NLO prediction is not statistically limited).
A similar study for jets measured by CMS at central rapidities can be found in Ref.~\cite{Rabbertz:1368241}.
Uncertainties from NP corrections and the renormalisation and factorisation scale variations, common to all
theoretical predictions, are added in quadrature and represented by the dashed (magenta) lines around the
ratio at unity in Figs.~\ref{fig:fwdjets_vs_PDFs}(b) and (c). Uncertainties on individual PDF sets are
displayed as bands. To improve the readability, the comparisons to data are performed separately in panel
(a) using the central values of all investigated PDF sets relative to CT10, in panel (b) for MSTW2008 and
NNPDF2.1, and in panel (c) for HERAPDF1.0 and ABKM09.

All NLO predictions for different PDF are similar and consistent with the data,
although they tend to systematically overestimate the central values of the measured forward jet
cross sections by $\sim$20\% in all $\pt$ bins. A similar overestimate 
has been observed for jets at more central pseudorapidities~\cite{Chatrchyan:2011me,Rabbertz:1368241}.

\subsection{Forward-central dijet spectra}

\begin{figure}[!Hhtbp]
   \centering
   \subfigure[]{\includegraphics[width=0.499\textwidth,height=7.cm]{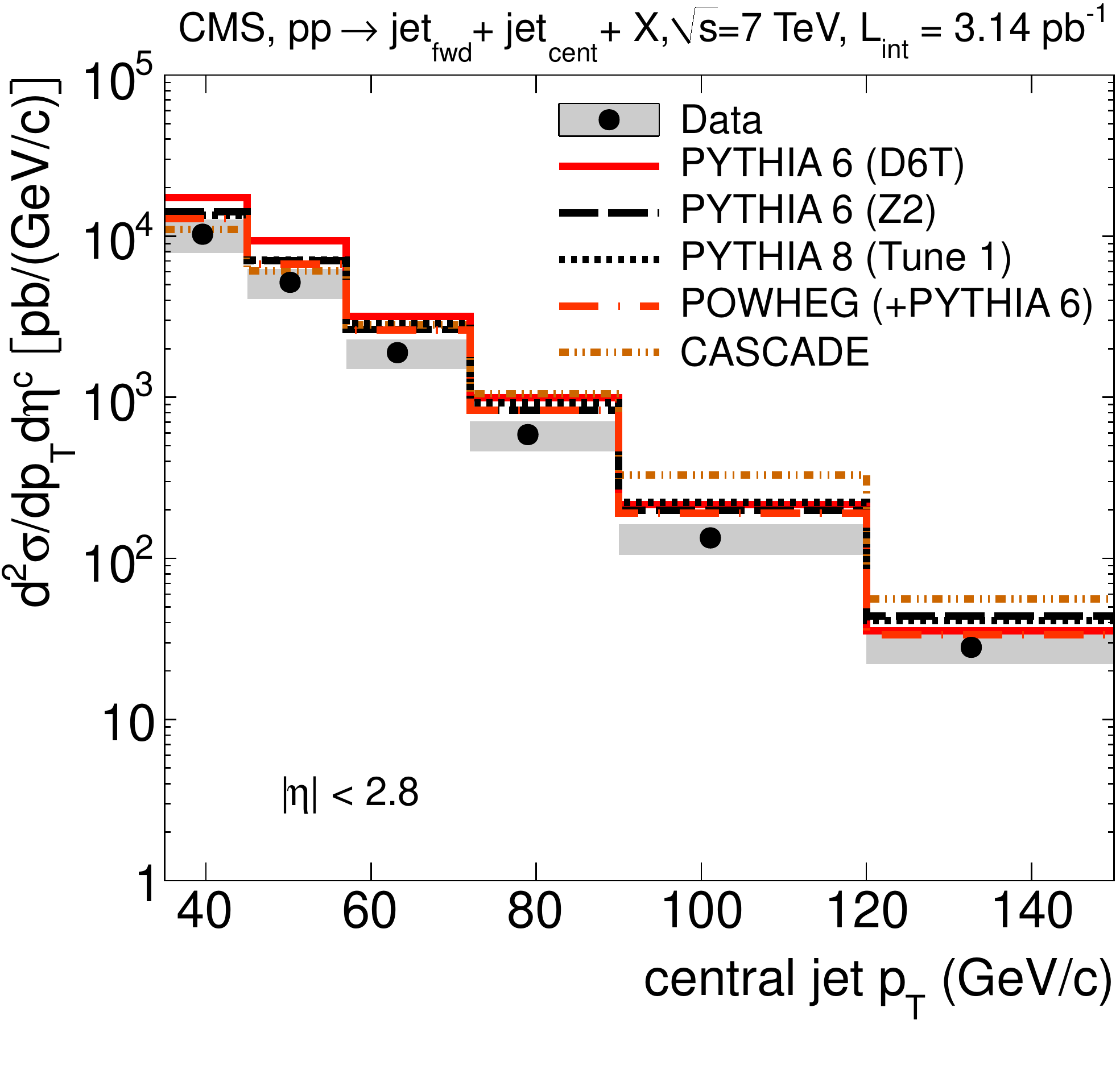}}
   \subfigure[]{\includegraphics[width=0.499\textwidth,height=7.cm]{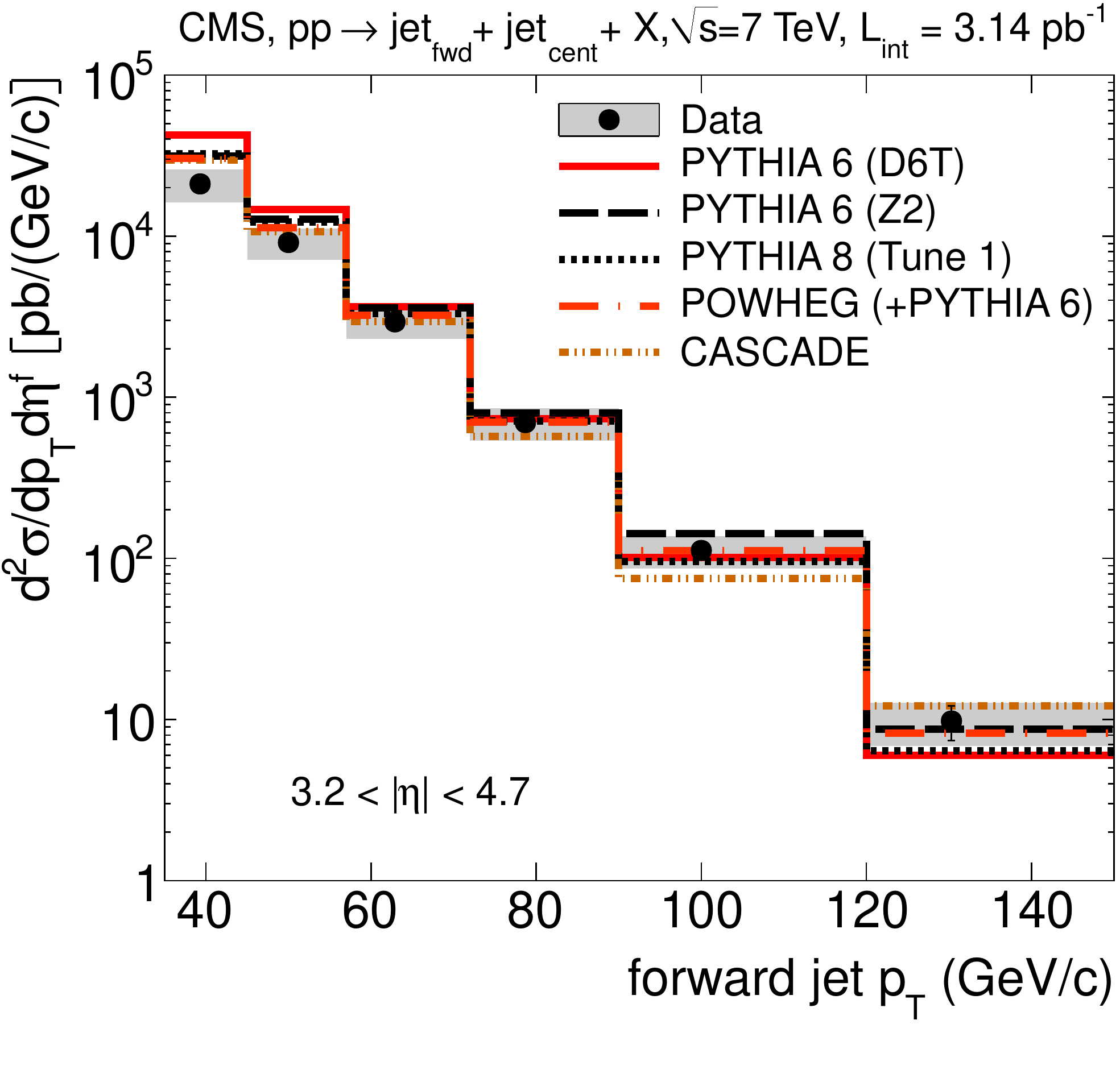}}\\

   \subfigure[]{\includegraphics[width=0.499\textwidth,height=7.cm]{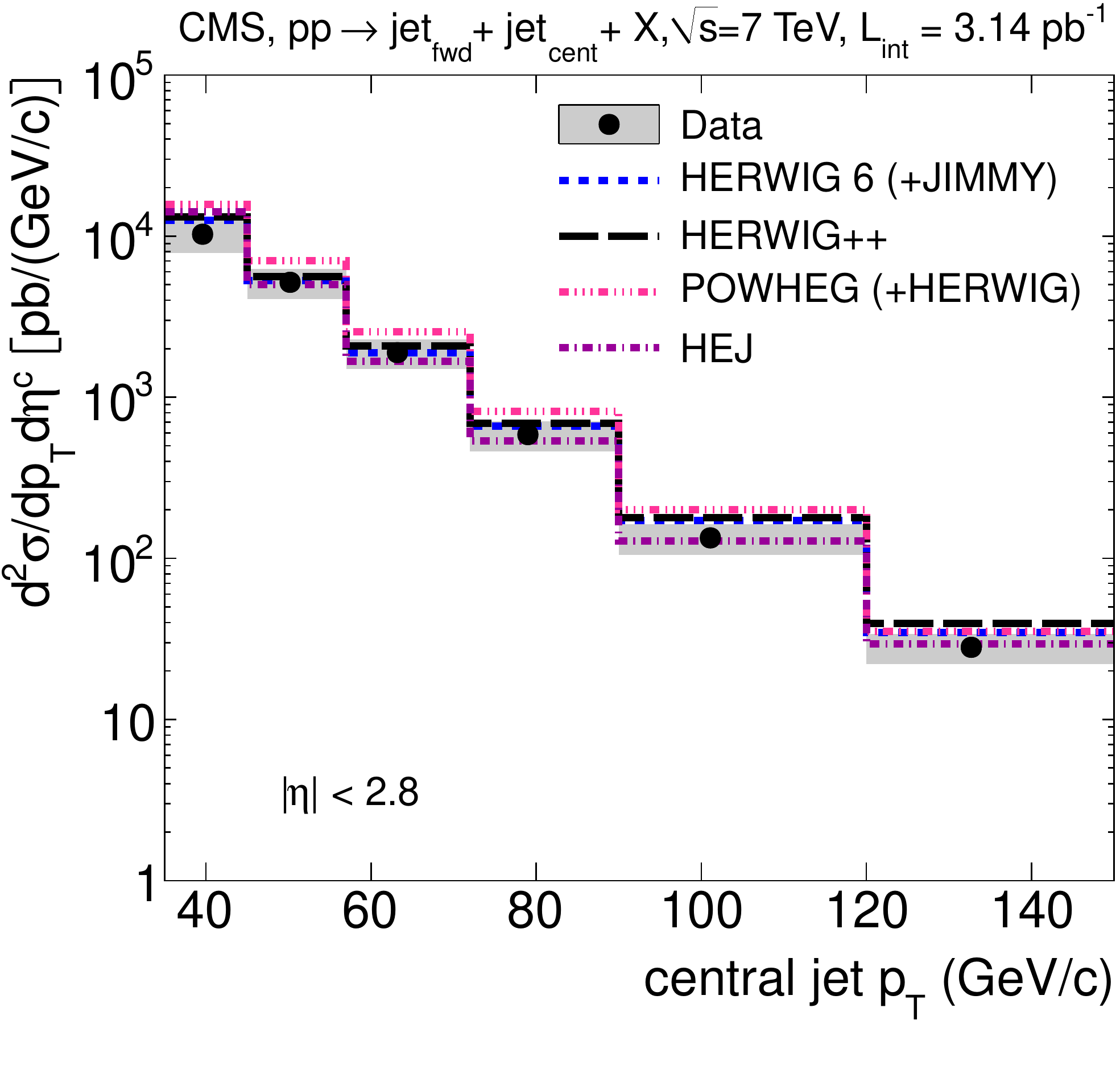}}
   \subfigure[]{\includegraphics[width=0.499\textwidth,height=7.cm]{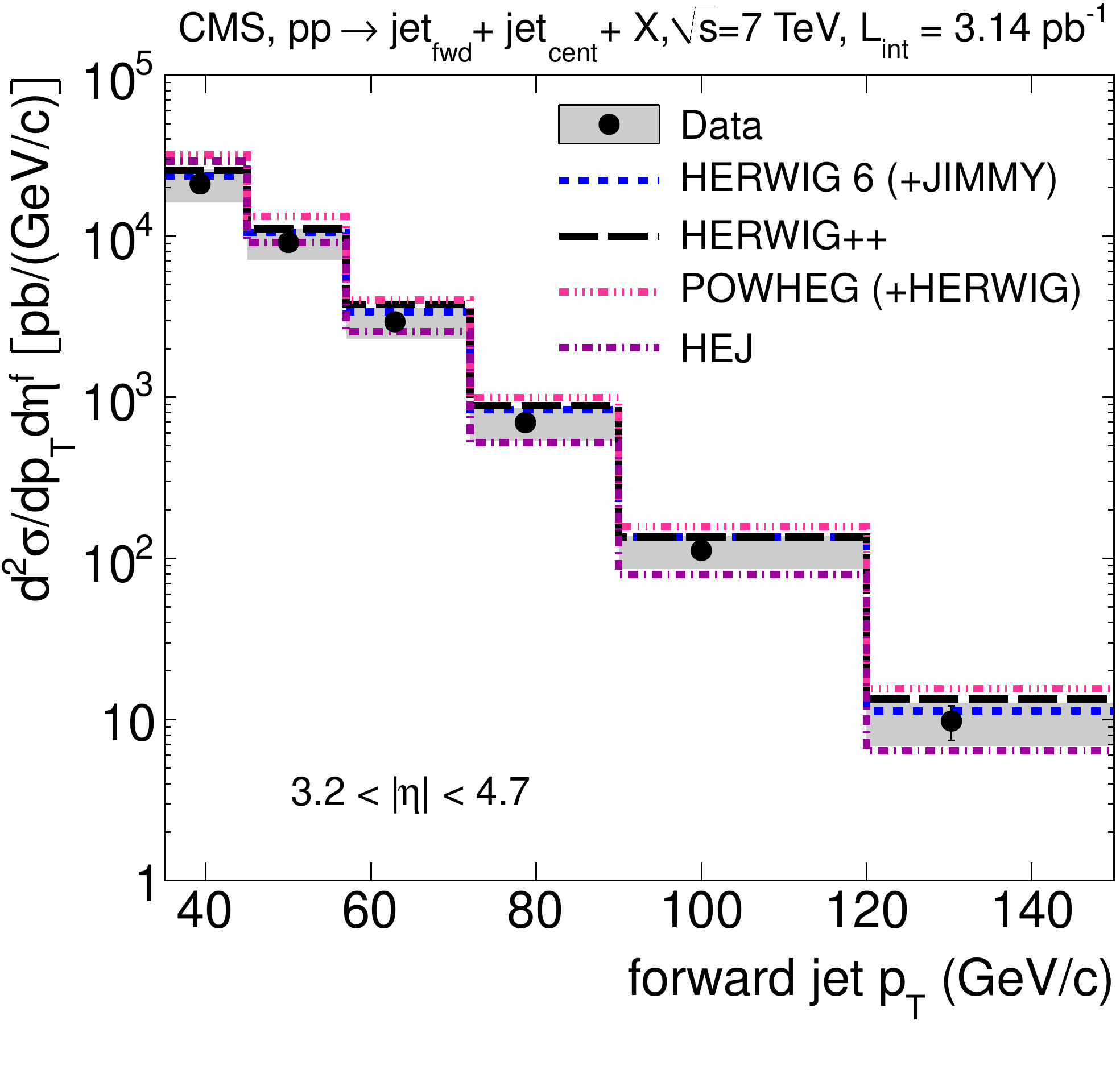}}
   \caption{Differential cross sections as a function of jet $\pt$ for dijet events with at least one central jet ((a) and
    (c)) and one forward jet ((b) and (d)), compared to predictions from several models. 
    The error bars on all data points (which, in (a) and (c), are smaller than the size of the markers)
    reflect just statistical uncertainties, with systematic uncertainties plotted as grey bands.}
   \label{fig:Xsec_had}
\end{figure}

The fully corrected $\pt$-dependence of the cross section for the simultaneous production of at least one forward
and at least one central jet is presented in Fig.~\ref{fig:Xsec_had} (a) and (c) for central
and in (b) and (d) for forward jets, respectively. The grey bands indicate the systematic uncertainties.
The cross sections obtained with \PYTHIA~6 (version 6.422) for D6T and Z2 tunes, \PYTHIA~8 (version 8.135),
\POWHEG\ (using \PYTHIA\ for parton showering and hadronisation), and \textsc{cascade} (version 2.2.04)
are superimposed on the data in panels (a) and (b), along with those for \HERWIG~6 (version 6.510.3),
\textsc{herwig++} (version 2.3), \POWHEG\ (using \HERWIG\ for parton showering and hadronisation),
and \textsc{hej}, shown in panels (c) and (d).

\begin{figure}[!Hhtbp]
   \centering
   \subfigure[]{\includegraphics[width=0.499\textwidth]{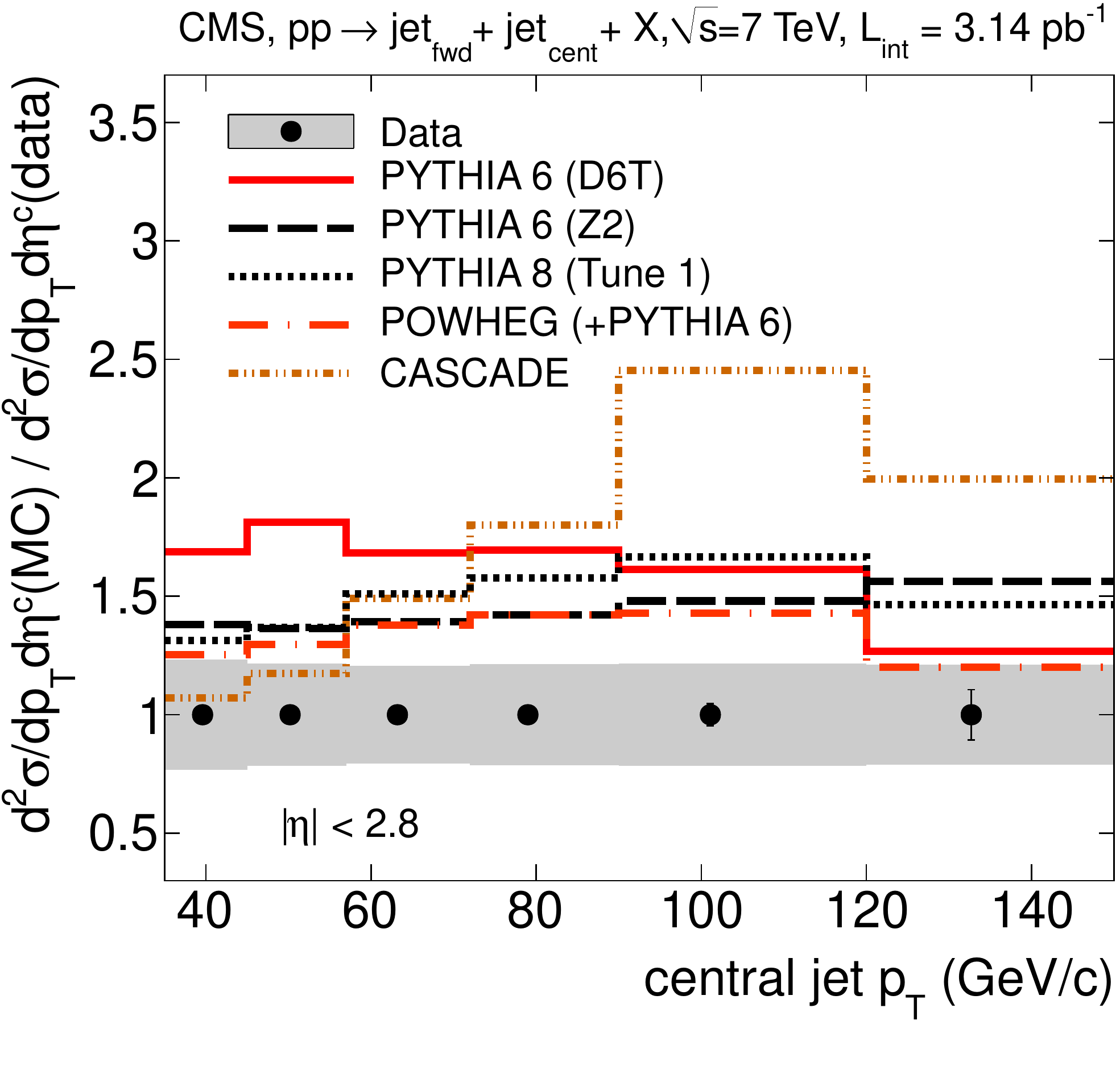}}
   \subfigure[]{\includegraphics[width=0.499\textwidth]{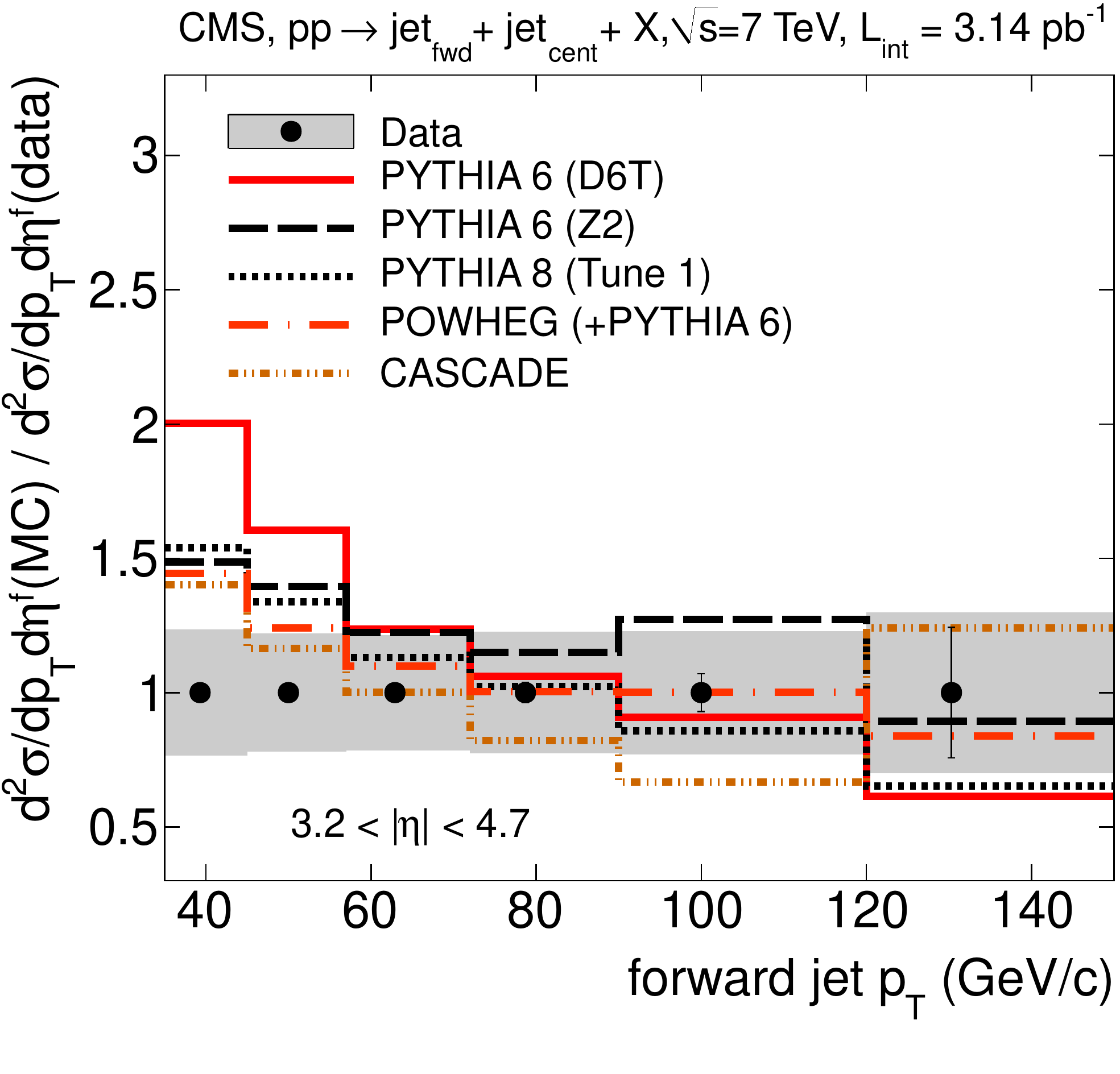}}\\

   \subfigure[]{\includegraphics[width=0.499\textwidth]{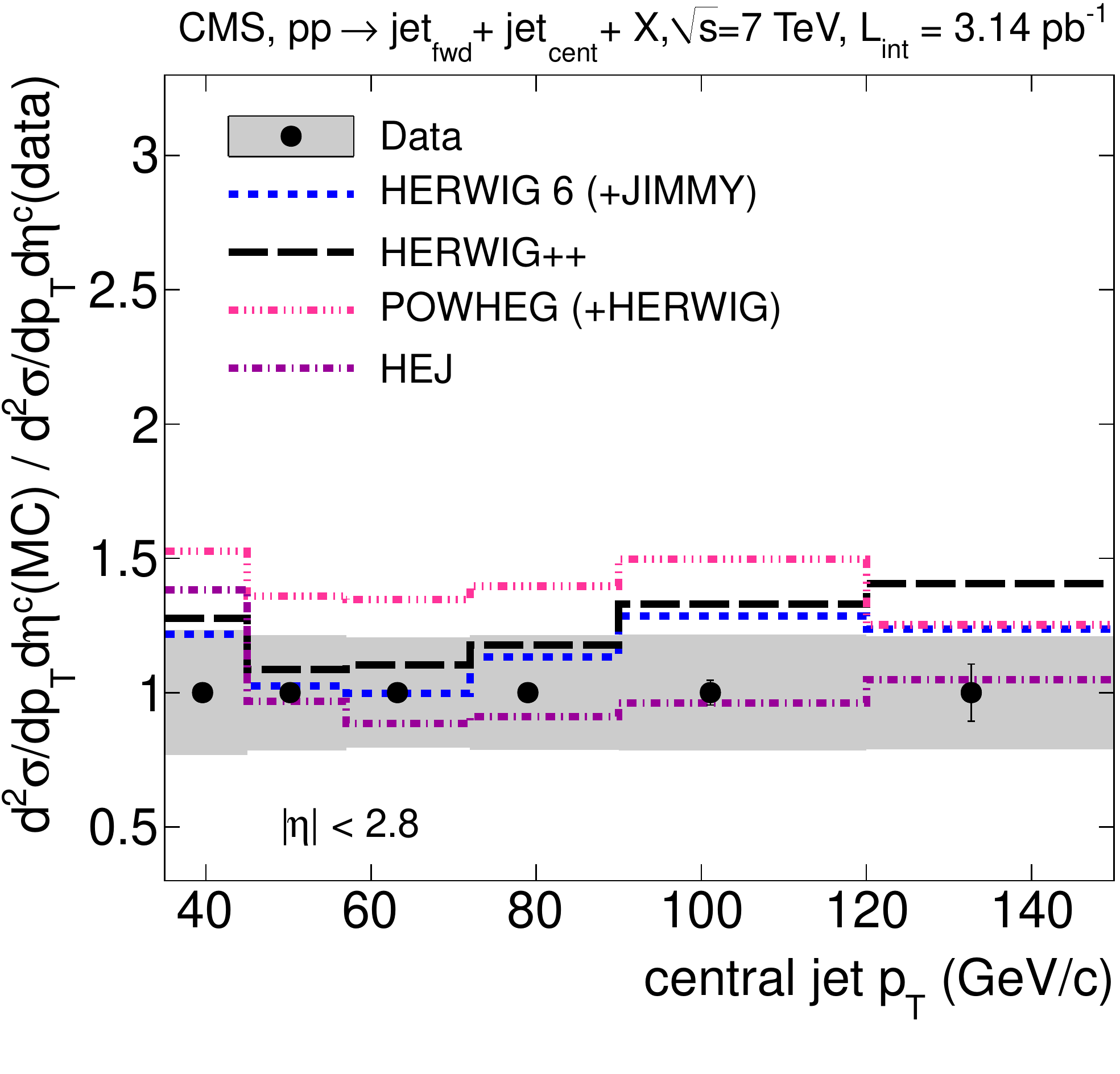}}
   \subfigure[]{\includegraphics[width=0.499\textwidth]{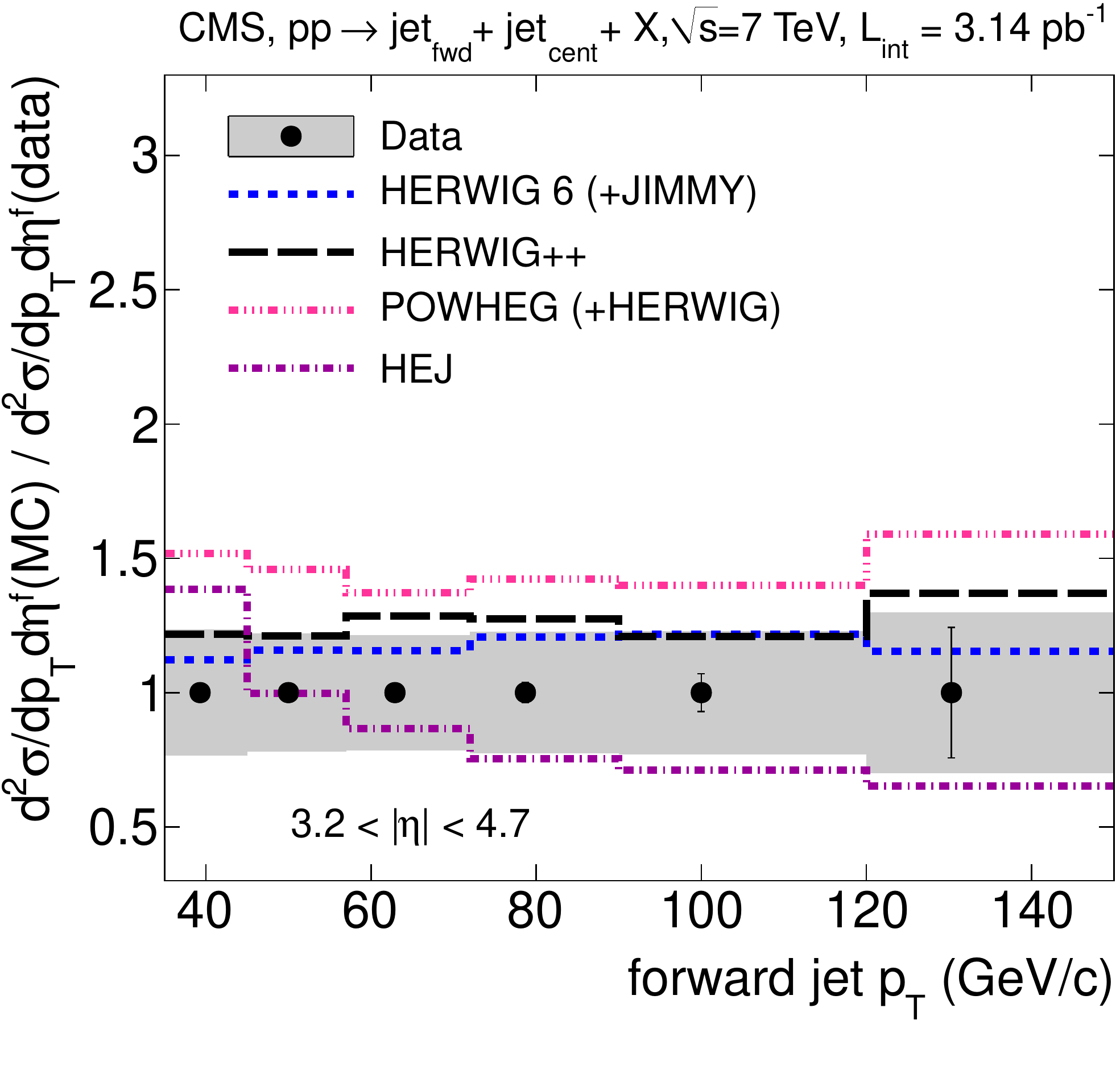}}
   \caption{Ratio of theory to data for differential cross sections as a function of
   $\pt$, for central ((a) and (c)) and forward ((b) and (d)) jets produced in
   dijet events. The error bars on all data points reflect just statistical uncertainties,
   with systematic uncertainties plotted as grey bands.}
   \label{fig:fwdJetPt_Xsec_ratio}
\end{figure}

The compatibility of the different models with the measured cross sections
is examined through the ratios of predictions to data
as a function of jet $\pt$ in Fig.~\ref{fig:fwdJetPt_Xsec_ratio}.
Most models tend to predict larger values than observed.
The \HERWIG\ and \textsc{herwig}++ MC event generators that use angular ordering for parton
showering appear to be consistent with the data. The other generators,
and different tunes, do not describe the data over the full range of $\pt$ values.
\PYTHIA~8 with Tune 1 and \PYTHIA~6 with Tune Z2 ($\pt$-ordered showering)
describe the data better than Tune D6T ($Q^2$-ordered showering).
The Z2 parameterisation tuned to the underlying event at the LHC, although reproducing
the central jet spectrum more satisfactorily than D6T or \PYTHIA~8, still lies well above the data
(the same holds true at lower $\pt$ for the forward-jet spectrum).
The discrepancy between \PYTHIA and data is therefore only partly reduced
through changes of the modelling of underlying event and initial- and final-state radiation.

The NLO MC \POWHEG\ matched to the \HERWIG\ parton shower reproduces
the dependence on $\pt$, but not the normalisation, which is overestimated by $\sim$40\%.
\textsc{cascade} predicts a different $\pt$-dependence which might come
from the initial-state parton showers~\cite{Deak:2010gk} which are
very sensitive to the unintegrated parton densities.
The \textsc{hej} code, used only at parton level here, describes the data reasonably well.

The discrepancies in the comparison of dijet data with MC models are larger for jets at
central values of $\eta$ in Figs.~\ref{fig:fwdJetPt_Xsec_ratio} (a) and (c).
In the case of forward jets, the comparison of the inclusive $\pt$ spectrum (Fig.~\ref{fig:xsection})
with that requiring the simultaneous presence of a jet in the central pseudorapidity region
(Figs.~\ref{fig:Xsec_had} (b) and (d)) 
shows that the inclusive spectrum is about a factor of four higher in the lowest $\pt$ bin but that
both distributions agree progressively better at larger $\pt$ values. This
suggests that inclusive forward jets of $\pt \approx$~35--70\GeVc may be
balanced by other forward jets or by soft central jets that do not surpass the
$\pt$ threshold of 35\GeVc, thereby producing the overall deficit of central jets
in the data shown in Figs.~\ref{fig:fwdJetPt_Xsec_ratio} (a) and (c).
These results confirm that the mechanisms for multijet production over large intervals in $\eta$
currently implemented in QCD models used at the LHC can be further constrained through
measurements of differential distributions like those presented here.
This work complements other studies based on jet multiplicities or on
$\pt$-integrated cross sections as a function of the jet $\Delta\eta$ separation.

\section{Summary}
\label{sec:Conclusions}

The inclusive production cross section for forward jets has been measured as a function of $\pt$,
in the pseudorapidity range 3.2~$<|\eta|<$~4.7. Also, the single-jet cross section has been measured
for the two leading jets in inclusive dijet events containing at least one forward and
one central jet (defined by the region $|\eta|<$~2.8).
The data are based on 3.14~pb$^{-1}$ of integrated luminosity collected
by the CMS detector in proton-proton collisions at $\sqrt{s}$~=~7~TeV.
Jets were reconstructed using the anti-$k_T$ algorithm ($R$~=~0.5) in the $\pt$ range 35--150\GeVc.
The total systematic uncertainties are $\pm$(20--30)\%, dominated by the absolute jet energy scale.
Within the current experimental and theoretical uncertainties, perturbative QCD calculations,
as implemented in the parton-shower event generators \PYTHIA\ and \HERWIG,
as well as in the combined DGLAP+BFKL resummation of the \textsc{cascade} model
and with the extra wide-angle gluon radiations included in the \textsc{hej} model,
are in good agreement with the measured inclusive single-jet forward cross section.
Calculations at NLO accuracy using recent sets of PDF also describe correctly the
$\pt$ dependence of the data, although the predicted absolute cross sections are about 20\% too large.

For the inclusive dijet events, all \PYTHIA\ tunes are found to overestimate the absolute cross
sections for the simultaneous production of jets above $\pt$~=~35\GeVc in the central and forward regions.
The agreement is poor for the entire central-jet spectrum and at smallest $\pt$ for forward jets.
The \HERWIG\ event generator provides a better description of both differential cross sections,
including their normalisations.
NLO contributions from \POWHEG\ to both of these parton-shower MC generators
enhance the cross sections at all $\pt$ and thereby the disagreement with data.
Calculations including resummation of low-$x$ logarithms, as in the \textsc{cascade}
Monte Carlo, do not reproduce the central-jet spectrum very well, but alternative approaches
that account for multijet BFKL-like topologies, such as in the \textsc{hej} model,
show reasonable agreement with the dijet data. 
The above measurements provide a valuable test of perturbative QCD in the forward region
of proton-proton collisions at the highest available energies, as well as a first check of
models for multijet production which are relevant to other processes at the LHC,
such as vector-boson fusion, characterised by forward/backward jet production.

\section*{Acknowledgments}

We wish to express our gratitude to Jeppe Andersen and Jenni Smillie for fruitful discussions
on the comparisons of the measurements with the \textsc{hej} model predictions,
and to Gavin Salam for useful exchanges on the theoretical calculations.

\hyphenation{Bundes-ministerium Forschungs-gemeinschaft Forschungs-zentren}
We wish to congratulate our colleagues in the CERN accelerator departments for the excellent performance of
the LHC machine. We thank the technical and administrative staff at CERN and other CMS institutes. This work
was supported by the Austrian Federal Ministry of Science and Research; the Belgium Fonds de la Recherche
Scientifique, and Fonds voor Wetenschappelijk Onderzoek; the Brazilian Funding Agencies (CNPq, CAPES, FAPERJ,
and FAPESP); the Bulgarian Ministry of Education and Science; CERN; the Chinese Academy of Sciences, Ministry
of Science and Technology, and National Natural Science Foundation of China; the Colombian Funding Agency
(COLCIENCIAS); the Croatian Ministry of Science, Education and Sport; the Research Promotion Foundation,
Cyprus; the Estonian Academy of Sciences and NICPB; the Academy of Finland, Finnish Ministry of Education and
Culture, and Helsinki Institute of Physics; the Institut National de Physique Nucl\'eaire et de Physique des
Particules~/~CNRS, and Commissariat \`a l'\'Energie Atomique et aux \'Energies Alternatives~/~CEA, France; the
Bundesministerium f\"ur Bildung und Forschung, Deutsche Forschungsgemeinschaft, and Helmholtz-Gemeinschaft
Deutscher Forschungszentren, Germany; the General Secretariat for Research and Technology, Greece; the
National Scientific Research Foundation, and National Office for Research and Technology, Hungary; the
Department of Atomic Energy and the Department of Science and Technology, India; the Institute for Studies in
Theoretical Physics and Mathematics, Iran; the Science Foundation, Ireland; the Istituto Nazionale di Fisica
Nucleare, Italy; the Korean Ministry of Education, Science and Technology and the World Class University
program of NRF, Korea; the Lithuanian Academy of Sciences; the Mexican Funding Agencies (CINVESTAV, CONACYT,
SEP, and UASLP-FAI); the Ministry of Science and Innovation, New Zealand; the Pakistan Atomic Energy
Commission; the Ministry of Science and Higher Education and the National Science Centre, Poland; the
Funda\c{c}\~ao para a Ci\^encia e a Tecnologia, Portugal; JINR (Armenia, Belarus, Georgia, Ukraine,
Uzbekistan); the Ministry of Education and Science of the Russian Federation, the Federal Agency of Atomic
Energy of the Russian Federation, Russian Academy of Sciences, and the Russian Foundation for Basic Research;
the Ministry of Science and Technological Development of Serbia; the Ministerio de Ciencia e Innovaci\'on, and
Programa Consolider-Ingenio 2010, Spain; the Swiss Funding Agencies (ETH Board, ETH Zurich, PSI, SNF, UniZH,
Canton Zurich, and SER); the National Science Council, Taipei; the Scientific and Technical Research Council
of Turkey, and Turkish Atomic Energy Authority; the Science and Technology Facilities Council, UK; the US
Department of Energy, and the US National Science Foundation.

Individuals have received support from the Marie-Curie programme and the European Research Council (European
Union); the Leventis Foundation; the A. P. Sloan Foundation; the Alexander von Humboldt Foundation; the
Belgian Federal Science Policy Office; the Fonds pour la Formation \`a la Recherche dans l'Industrie et dans
l'Agriculture (FRIA-Belgium); the Agentschap voor Innovatie door Wetenschap en Technologie (IWT-Belgium); the
Council of Science and Industrial Research, India; and the HOMING PLUS programme of Foundation for Polish
Science, cofinanced from European Union, Regional Development Fund.

\bibliography{auto_generated}  

\providecommand{\href}[2]{#2}\begingroup\raggedright\begin{thebibliography}{10}%
\makeatletter
\providecommand{\hrefCMSnoop }[0]{\@secondoftwo}%
\makeatother
\providecommand{\doi}{\texttt{doi:}\begingroup \urlstyle{tt}\Url}

\bibitem{Ellis:2007ib}
\hrefCMSnoop {} {S.~Ellis, J.~Huston, K.~Hatakeyama{ et~al.}, ``{Jets in
  hadron-hadron collisions}'',} \textit{ Prog. Part. Nucl. Phys.} \textbf{ 60}
  (2008) 484,
  \href{http://dx.doi.org/10.1016/j.ppnp.2007.12.002}{\doi{10.1016/j.ppnp.2007.12.002}},
  \href{http://www.arXiv.org/abs/0712.2447}{\texttt{ arXiv:0712.2447}}.

\bibitem{Chatrchyan:2011me}
\hrefCMSnoop {} {{ CMS} Collaboration, ``{Measurement of the Inclusive Jet
  Cross Section in pp Collisions at sqrt(s) = 7 TeV}'',} \textit{ Phys. Rev.
  Lett.} \textbf{ 107} (2011) 132001,
  \href{http://dx.doi.org/10.1103/PhysRevLett.107.132001}{\doi{10.1103/PhysRevLett.107.132001}},
\href{http://www.arXiv.org/abs/1106.0208}{\texttt{ arXiv:1106.0208}}.

\bibitem{Aad:2011fc}
\hrefCMSnoop {} {{ ATLAS Collaboration} Collaboration, ``{Measurement of
  inclusive jet and dijet production in pp collisions at $\sqrt(s) = 7$ TeV
  using the ATLAS detector}'',} (2011).
  \href{http://www.arXiv.org/abs/1112.6297}{\texttt{ arXiv:1112.6297}}.
Submitted to \textit{Phys. Rev. D}.

\bibitem{Abazov:2008hua}
\hrefCMSnoop {} {{ D0} Collaboration, ``{Measurement of the inclusive jet
  cross-section in $p \bar{p}$ collisions at $\sqrt{s}^{1/2}$ =1.96 TeV}'',}
  \textit{ Phys. Rev. Lett.} \textbf{ 101} (2008) 062001,
  \href{http://dx.doi.org/10.1103/PhysRevLett.101.062001}{\doi{10.1103/PhysRevLett.101.062001}},
\href{http://www.arXiv.org/abs/0802.2400}{\texttt{ arXiv:0802.2400}}.

\bibitem{Aaltonen:2008eq}
\hrefCMSnoop {} {{ CDF} Collaboration, ``{Measurement of the Inclusive Jet
  Cross Section at the Fermilab Tevatron p-pbar Collider Using a Cone-Based Jet
  Algorithm}'',} \textit{ Phys. Rev. D} \textbf{ 78} (2008) 052006,
  \href{http://dx.doi.org/10.1103/PhysRevD.78.052006}{\doi{10.1103/PhysRevD.78.052006}},
\href{http://www.arXiv.org/abs/0807.2204}{\texttt{ arXiv:0807.2204}}.

\bibitem{HERAPDF10:2009wt}
\hrefCMSnoop {} {{ H1 and ZEUS} Collaboration, ``{Combined Measurement and QCD
  Analysis of the Inclusive ep Scattering Cross Sections at HERA}'',} \textit{
  JHEP} \textbf{ 01} (2010) 109,
  \href{http://dx.doi.org/10.1007/JHEP01(2010)109}{\doi{10.1007/JHEP01(2010)109}},
\href{http://www.arXiv.org/abs/0911.0884}{\texttt{ arXiv:0911.0884}}.

\bibitem{Cerci:2008xv}
\hrefCMSnoop {} {S.~Cerci and D.~d'Enterria, ``{Low-x QCD studies with forward
  jets in proton-proton collisions at $\sqrt{s}$ = 14 TeV in CMS}'',} \textit{
  AIP Conf. Proc.} \textbf{ 1105} (2009) 28,
  \href{http://dx.doi.org/10.1063/1.3122196}{\doi{10.1063/1.3122196}},
\href{http://www.arXiv.org/abs/0812.2665}{\texttt{ arXiv:0812.2665}}.

\bibitem{Gribov:1972ri}
\hrefCMSnoop {} {V.~N. Gribov and L.~N. Lipatov, ``{Deep inelastic {\it ep}
  scattering in perturbation theory}'',} \textit{ Sov. J. Nucl. Phys.} \textbf{
  15} (1972)
438.

\bibitem{Lipatov:1974qm}
\hrefCMSnoop {} {L.~N. Lipatov, ``{The parton model and perturbation
  theory}'',} \textit{ Sov. J. Nucl. Phys.} \textbf{ 20} (1975)
94.

\bibitem{Altarelli:1977zs}
\hrefCMSnoop {} {G.~Altarelli and G.~Parisi, ``{Asymptotic Freedom in Parton
  Language}'',} \textit{ Nucl. Phys. B} \textbf{ 126} (1977) 298,
\href{http://dx.doi.org/10.1016/0550-3213(77)90384-4}{\doi{10.1016/0550-3213(77)90384-4}}.

\bibitem{Dokshitzer:1977sg}
\hrefCMSnoop {} {Y.~L. Dokshitzer, ``{Calculation of the Structure Functions
  for Deep Inelastic Scattering and {\it e+e-} Annihilation by Perturbation
  Theory in Quantum Chromodynamics}'',} \textit{ Sov. Phys. JETP} \textbf{ 46}
  (1977)
641.

\bibitem{Kuraev:1976ge}
\hrefCMSnoop {} {E.~A. Kuraev, L.~N. Lipatov, and V.~S. Fadin, ``{Multi -
  Reggeon Processes in the Yang-Mills Theory}'',} \textit{ Sov. Phys. JETP}
  \textbf{ 44} (1976)
443.

\bibitem{Kuraev:1977fs}
\hrefCMSnoop {} {E.~A. Kuraev, L.~N. Lipatov, and V.~S. Fadin, ``{The
  Pomeranchuk Singularity in Nonabelian Gauge Theories}'',} \textit{ Sov. Phys.
  JETP} \textbf{ 45} (1977)
199.

\bibitem{Balitsky:1978ic}
\hrefCMSnoop {} {I.~I. Balitsky and L.~N. Lipatov, ``{The Pomeranchuk
  Singularity in Quantum Chromodynamics}'',} \textit{ Sov. J. Nucl. Phys.}
  \textbf{ 28} (1978)
822.

\bibitem{Ciafaloni:1987ur}
\hrefCMSnoop {} {M.~Ciafaloni, ``{Coherence Effects in Initial Jets at Small
  $q^2$/$s$}'',} \textit{ Nucl. Phys. B} \textbf{ 296} (1988) 49,
\href{http://dx.doi.org/10.1016/0550-3213(88)90380-X}{\doi{10.1016/0550-3213(88)90380-X}}.

\bibitem{Catani:1989yc}
\hrefCMSnoop {} {S.~Catani, F.~Fiorani, and G.~Marchesini, ``{QCD Coherence in
  Initial State Radiation}'',} \textit{ Phys. Lett. B} \textbf{ 234} (1990)
  339,
\href{http://dx.doi.org/10.1016/0370-2693(90)91938-8}{\doi{10.1016/0370-2693(90)91938-8}}.

\bibitem{Catani:1989sg}
\hrefCMSnoop {} {S.~Catani, F.~Fiorani, and G.~Marchesini, ``{Small $x$
  Behavior of Initial State Radiation in Perturbative QCD}'',} \textit{ Nucl.
  Phys. B} \textbf{ 336} (1990) 18,
\href{http://dx.doi.org/10.1016/0550-3213(90)90342-B}{\doi{10.1016/0550-3213(90)90342-B}}.

\bibitem{Marchesini:1994wr}
\hrefCMSnoop {} {G.~Marchesini, ``{QCD coherence in the structure function and
  associated distributions at small x}'',} \textit{ Nucl. Phys. B} \textbf{
  445} (1995) 49,
  \href{http://dx.doi.org/10.1016/0550-3213(95)00149-M}{\doi{10.1016/0550-3213(95)00149-M}},
\href{http://www.arXiv.org/abs/hep-ph/9412327}{\texttt{ arXiv:hep-ph/9412327}}.

\bibitem{Gelis:2010nm}
\hrefCMSnoop {} {F.~Gelis, E.~Iancu, J.~Jalilian-Marian{ et~al.}, ``{The Color
  Glass Condensate}'',} \textit{ Ann. Rev. Nucl. Part. Sci.} \textbf{ 60}
  (2010) 463,
  \href{http://dx.doi.org/10.1146/annurev.nucl.010909.083629}{\doi{10.1146/annurev.nucl.010909.083629}},
\href{http://www.arXiv.org/abs/1002.0333}{\texttt{ arXiv:1002.0333}}.

\bibitem{Figy:2003nv}
\hrefCMSnoop {} {T.~Figy, C.~Oleari, and D.~Zeppenfeld, ``{Next-to-leading
  order jet distributions for Higgs boson production via weak-boson fusion}'',}
  \textit{ Phys. Rev. D} \textbf{ 68} (2003) 073005,
  \href{http://dx.doi.org/10.1103/PhysRevD.68.073005}{\doi{10.1103/PhysRevD.68.073005}},
\href{http://www.arXiv.org/abs/hep-ph/0306109}{\texttt{ arXiv:hep-ph/0306109}}.

\bibitem{Berger:2010xi}
\hrefCMSnoop {} {C.~F. Berger, C.~Marcantonini, I.~W. Stewart{ et~al.},
  ``{Higgs Production with a Central Jet Veto at NNLL+NNLO}'',} \textit{ JHEP}
  \textbf{ 04} (2011) 092,
  \href{http://dx.doi.org/10.1007/JHEP04(2011)092}{\doi{10.1007/JHEP04(2011)092}},
\href{http://www.arXiv.org/abs/1012.4480}{\texttt{ arXiv:1012.4480}}.

\bibitem{Butterworth:2002tt}
\hrefCMSnoop {} {J.~Butterworth, B.~Cox, and J.~R. Forshaw, ``{$W W$ scattering
  at the CERN LHC}'',} \textit{ Phys. Rev. D} \textbf{ 65} (2002) 096014,
  \href{http://dx.doi.org/10.1103/PhysRevD.65.096014}{\doi{10.1103/PhysRevD.65.096014}},
\href{http://www.arXiv.org/abs/hep-ph/0201098}{\texttt{ arXiv:hep-ph/0201098}}.

\bibitem{Aad:2011jz}
\hrefCMSnoop {} {{ ATLAS} Collaboration, ``{Measurement of dijet production
  with a veto on additional central jet activity in pp collisions at
  $\sqrt{s}$~=~7 TeV using the ATLAS detector}'',} \textit{ JHEP} \textbf{ 09}
  (2011) 053,
  \href{http://dx.doi.org/10.1007/JHEP09(2011)053}{\doi{10.1007/JHEP09(2011)053}},
\href{http://www.arXiv.org/abs/1107.1641}{\texttt{ arXiv:1107.1641}}.

\bibitem{pythia}
\hrefCMSnoop {} {T.~Sj{\"{o}}strand, S.~Mrenna, and P.~Skands, ``{PYTHIA 6.4
  Physics and Manual}'',} \textit{ JHEP} \textbf{ 05} (2006) 026,
\href{http://www.arXiv.org/abs/hep-ph/0603175}{\texttt{ arXiv:hep-ph/0603175}}.

\bibitem{Sjostrand:2007gs}
\hrefCMSnoop {} {T.~Sj{\"{o}}strand, S.~Mrenna, and P.~Skands, ``{A Brief
  Introduction to PYTHIA 8.1}'',} \textit{ Comput. Phys. Commun.} \textbf{ 178}
  (2008) 852,
  \href{http://dx.doi.org/10.1016/j.cpc.2008.01.036}{\doi{10.1016/j.cpc.2008.01.036}},
  \href{http://www.arXiv.org/abs/0710.3820}{\texttt{ arXiv:0710.3820}}.

\bibitem{herwig6}
\hrefCMSnoop {} {G.~Marchesini {et~al.}, ``{HERWIG: A Monte Carlo event
  generator for simulating hadron emission reactions with interfering gluons.
  Version 5.1 - April 1991}'',} \textit{ Comput. Phys. Commun.} \textbf{ 67}
  (1992) 465,
\href{http://dx.doi.org/10.1016/0010-4655(92)90055-4}{\doi{10.1016/0010-4655(92)90055-4}}.

\bibitem{jimmy}
\hrefCMSnoop {} {J.~M. Butterworth, J.~R. Forshaw, and M.~H. Seymour,
  ``{Multiparton interactions in photoproduction at HERA}'',} \textit{ Z. Phys.
  C} \textbf{ 72} (1996) 637,
  \href{http://dx.doi.org/10.1007/s002880050286}{\doi{10.1007/s002880050286}},
\href{http://www.arXiv.org/abs/hep-ph/9601371}{\texttt{ arXiv:hep-ph/9601371}}.

\bibitem{Bahr:2008tx}
\hrefCMSnoop {} {M.~Bahr, S.~Gieseke, M.~Gigg{ et~al.}, ``{Herwig++ 2.2 Release
  Note}'',}
\href{http://www.arXiv.org/abs/0804.3053}{\texttt{ arXiv:0804.3053}}.

\bibitem{Nagy:2001fj}
\hrefCMSnoop {} {Z.~Nagy, ``{Three-jet cross sections in hadron hadron
  collisions at next-to-leading order}'',} \textit{ Phys. Rev. Lett.} \textbf{
  88} (2002) 122003,
  \href{http://dx.doi.org/10.1103/PhysRevLett.88.122003}{\doi{10.1103/PhysRevLett.88.122003}},
\href{http://www.arXiv.org/abs/hep-ph/0110315}{\texttt{ arXiv:hep-ph/0110315}}.

\bibitem{Nagy:2003tz}
\hrefCMSnoop {} {Z.~Nagy, ``{Next-to-leading order calculation of three jet
  observables in hadron hadron collision}'',} \textit{ Phys. Rev. D} \textbf{
  68} (2003) 094002,
  \href{http://dx.doi.org/10.1103/PhysRevD.68.094002}{\doi{10.1103/PhysRevD.68.094002}},
\href{http://www.arXiv.org/abs/hep-ph/0307268}{\texttt{ arXiv:hep-ph/0307268}}.

\bibitem{Frixione:2007vw}
\hrefCMSnoop {} {{S. Frixione, P. Nason and C. Oleari}, ``{Matching NLO QCD
  computations with Parton Shower simulations: the POWHEG method}'',} \textit{
  JHEP} \textbf{ 11} (2007) 070,
  \href{http://dx.doi.org/10.1088/1126-6708/2007/11/070}{\doi{10.1088/1126-6708/2007/11/070}},
\href{http://www.arXiv.org/abs/0709.2092}{\texttt{ arXiv:0709.2092}}.

\bibitem{Jung:2000hk}
\hrefCMSnoop {} {H.~Jung and G.~P. Salam, ``{Hadronic final state predictions
  from CCFM: The hadron- level Monte Carlo generator CASCADE}'',} \textit{ Eur.
  Phys. J.} \textbf{ C19} (2001) 351,
  \href{http://dx.doi.org/10.1007/s100520100604}{\doi{10.1007/s100520100604}},
\href{http://www.arXiv.org/abs/hep-ph/0012143}{\texttt{ arXiv:hep-ph/0012143}}.

\bibitem{Jung:2010si}
\hrefCMSnoop {} {H.~Jung, S.~Baranov, M.~Deak{ et~al.}, ``{The CCFM Monte Carlo
  generator CASCADE 2.2.0}'',} \textit{ Eur. Phys. J.} \textbf{ C79} (2010)
  1237,
\href{http://www.arXiv.org/abs/1008.0152}{\texttt{ arXiv:1008.0152}}.

\bibitem{Andersen:2009nu}
\hrefCMSnoop {} {J.~R. Andersen and J.~M. Smillie, ``{Constructing All-Order
  Corrections to Multi-Jet Rates}'',} \textit{ JHEP} \textbf{ 01} (2010) 039,
  \href{http://dx.doi.org/10.1007/JHEP01(2010)039}{\doi{10.1007/JHEP01(2010)039}},
\href{http://www.arXiv.org/abs/0908.2786}{\texttt{ arXiv:0908.2786}}.

\bibitem{Andersen:2011hs}
\hrefCMSnoop {} {J.~R. Andersen and J.~M. Smillie, ``{Multiple Jets at the LHC
  with High Energy Jets}'',} \textit{ JHEP} \textbf{ 06} (2011) 010,
  \href{http://dx.doi.org/10.1007/JHEP06(2011)010}{\doi{10.1007/JHEP06(2011)010}},
\href{http://www.arXiv.org/abs/1101.5394}{\texttt{ arXiv:1101.5394}}.

\bibitem{Adolphi:2008zzk}
\hrefCMSnoop {} {{ CMS} Collaboration, ``The CMS experiment at the CERN LHC'',}
  \textit{ JINST} \textbf{ 03} (2008) S08004,
\href{http://dx.doi.org/10.1088/1748-0221/3/08/S08004}{\doi{10.1088/1748-0221/3/08/S08004}}.

\bibitem{Chatrchyan:2009qm}
\hrefCMSnoop {} {{ CMS} Collaboration, ``{Performance and Operation of the CMS
  Electromagnetic Calorimeter}'',} \textit{ JINST} \textbf{ 05} (2010) T03010,
  \href{http://dx.doi.org/10.1088/1748-0221/5/03/T03010}{\doi{10.1088/1748-0221/5/03/T03010}},
  \href{http://www.arXiv.org/abs/0910.3423}{\texttt{ arXiv:0910.3423}}.

\bibitem{Chatrchyan:2009vn}
\hrefCMSnoop {} {{ CMS} Collaboration, ``{Performance of the CMS Hadron
  Calorimeter with Cosmic Ray Muons and LHC Beam Data}'',} \textit{ JINST}
  \textbf{ 05} (2010) T03012,
  \href{http://dx.doi.org/10.1088/1748-0221/5/03/T03012}{\doi{10.1088/1748-0221/5/03/T03012}},
  \href{http://www.arXiv.org/abs/0911.4991}{\texttt{ arXiv:0911.4991}}.

\bibitem{HF}
\hrefCMSnoop {} {G.~Bayatian {et~al.}, ``{Design, performance and calibration
  of the CMS forward calorimeter wedges}'',} \textit{ Eur. Phys. J. C} \textbf{
  53} (2008) 139,
\href{http://dx.doi.org/10.1140/epjc/s10052-007-0459-4}{\doi{10.1140/epjc/s10052-007-0459-4}}.

\bibitem{Bayatian:2006zz}
\hrefCMSnoop {} {{ CMS} Collaboration, ``CMS Technical Design Report, Vol I'',}
  \textit{ Technical Report CERN-LHCC-2006-001$/$CMS-TDR-008-1} (2006)
Section 11.6.3.

\bibitem{CMS-PAS-JME-07-003}
\hrefCMSnoop {} {{ CMS} Collaboration, ``Performance of Jet Algorithms in
  CMS'',} \textit{ CMS PAS} \textbf{
  \href{http://cdsweb.cern.ch/record/1198227/files/JME-07-003-pas.pdf}{JME-07-003}}
  (2007).

\bibitem{track}
\hrefCMSnoop {} {{CMS Collaboration}, ``Tracking and Vertexing Results from
  First Collisions'',} \textit{ CMS PAS} \textbf{ TRK-10-001} (2010).

\bibitem{a-ktalg}
\hrefCMSnoop {} {M.~Cacciari, G.~P. Salam, and G.~Soyez, ``{The anti-kt jet
  clustering algorithm}'',} \textit{ JHEP} \textbf{ 04} (2008) 063,
  \href{http://dx.doi.org/10.1088/1126-6708/2008/04/063}{\doi{10.1088/1126-6708/2008/04/063}},
\href{http://www.arXiv.org/abs/0802.1189}{\texttt{ arXiv:0802.1189}}.

\bibitem{Cacciari:2011ma}
\hrefCMSnoop {} {M.~Cacciari, G.~P. Salam, and G.~Soyez, ``{FastJet user
  manual}'',} (2011).
\href{http://www.arXiv.org/abs/1111.6097}{\texttt{ arXiv:1111.6097}}.

\bibitem{Chatrchyan:2011ds}
\hrefCMSnoop {} {{ CMS} Collaboration, ``{Determination of Jet Energy
  Calibration and Transverse Momentum Resolution in CMS}'',} \textit{ JINST}
  \textbf{ 6} (2011) P11002,
  \href{http://dx.doi.org/10.1088/1748-0221/6/11/P11002}{\doi{10.1088/1748-0221/6/11/P11002}},
\href{http://www.arXiv.org/abs/1107.4277}{\texttt{ arXiv:1107.4277}}.

\bibitem{Agostinelli:2002hh}
\hrefCMSnoop {} {{ GEANT4} Collaboration, ``{GEANT4: A simulation toolkit}'',}
  \textit{ Nucl. Instrum. Meth. A} \textbf{ 506} (2003)
250--303.

\bibitem{arXiv:1105.1160}
\hrefCMSnoop {} {T.~Adye, ``Unfolding algorithms and tests using RooUnfold'',}
  (2011).
\href{http://www.arXiv.org/abs/1105.1160}{\texttt{ arXiv:1105.1160}}.

\bibitem{ansatz-one}
\hrefCMSnoop {} {S.~M. Berman, J.~D. Bjorken, and J.~B. Kogut, ``{Inclusive
  Processes at High Transverse Momentum}'',} \textit{ Phys. Rev. D} \textbf{ 4}
  (1971) 3388,
\href{http://dx.doi.org/10.1103/PhysRevD.4.3388}{\doi{10.1103/PhysRevD.4.3388}}.

\bibitem{ansatz-two}
\hrefCMSnoop {} {R.~Feynman, R.~Field, and G.~Fox, ``{Quantum-chromodynamic
  approach for the large-transverse-momentum production of particles and
  jets}'',} \textit{ Phys. Rev. D} \textbf{ 18} (1978) 3320.

\bibitem{CMS-DP-2011-002}
\href {http://cdsweb.cern.ch/record/1335668} {{ CMS} Collaboration, ``Absolute
  luminosity normalization'',} CMS Detector Performance note DP-2011-002, (Mar,
  2011).

\bibitem{tune}
\hrefCMSnoop {} {R.~Field, ``{Early LHC Underlying Event Data - Findings and
  Surprises}'',}
\href{http://www.arXiv.org/abs/1010.3558}{\texttt{ arXiv:1010.3558}}.

\bibitem{Alioli:2010xa}
\hrefCMSnoop {} {S.~Alioli, K.~Hamilton, P.~Nason{ et~al.}, ``{Jet pair
  production in POWHEG}'',} \textit{ JHEP} \textbf{ 04} (2011) 081,
  \href{http://dx.doi.org/10.1007/JHEP04(2011)081}{\doi{10.1007/JHEP04(2011)081}},
\href{http://www.arXiv.org/abs/1012.3380}{\texttt{ arXiv:1012.3380}}.

\bibitem{Kluge:2006xs}
\hrefCMSnoop {} {T.~Kluge, K.~Rabbertz, and M.~Wobisch, ``Fast pQCD
  calculations for PDF fits'',} in \textit{ {14th International Workshop on
  Deep Inelastic Scattering (DIS 2006), 20-24 Apr 2006}}, p.~483.
\newblock Tsukuba, Japan, April, 2006.
\newblock \href{http://www.arXiv.org/abs/hep-ph/0609285}{\texttt{
  arXiv:hep-ph/0609285}}.
\newblock
\href{http://dx.doi.org/10.1142/9789812706706_0110}{\doi{10.1142/9789812706706_0110}}.

\bibitem{Deak:2009xt}
\hrefCMSnoop {} {M.~Deak, F.~Hautmann, H.~Jung{ et~al.}, ``{Forward Jet
  Production at the Large Hadron Collider}'',} \textit{ JHEP} \textbf{ 09}
  (2009) 121,
  \href{http://dx.doi.org/10.1088/1126-6708/2009/09/121}{\doi{10.1088/1126-6708/2009/09/121}},
  \href{http://www.arXiv.org/abs/0908.0538}{\texttt{ arXiv:0908.0538}}.

\bibitem{Deak:2010gk}
\hrefCMSnoop {} {M.~Deak, F.~Hautmann, H.~Jung{ et~al.}, ``{Forward-Central Jet
  Correlations at the Large Hadron Collider}'',} (2010).
\href{http://www.arXiv.org/abs/1012.6037}{\texttt{ arXiv:1012.6037}}.

\bibitem{Pumplin:2002vw}
\hrefCMSnoop {} {J.~Pumplin, D.~Stump, J.~Huston{ et~al.}, ``{New generation of
  parton distributions with uncertainties from global QCD analysis}'',}
  \textit{ JHEP} \textbf{ 07} (2002) 012,
  \href{http://dx.doi.org/10.1088/1126-6708/2002/07/012}{\doi{10.1088/1126-6708/2002/07/012}},
\href{http://www.arXiv.org/abs/hep-ph/0201195}{\texttt{ arXiv:hep-ph/0201195}}.

\bibitem{Lai:1999wy}
\hrefCMSnoop {} {{ CTEQ} Collaboration, ``Global {QCD} analysis of parton
  structure of the nucleon: CTEQ5 parton distributions'',} \textit{ Eur. Phys.
  J. C} \textbf{ 12} (2000) 375--392,
  \href{http://dx.doi.org/10.1007/s100529900196}{\doi{10.1007/s100529900196}},
\href{http://www.arXiv.org/abs/hep-ph/9903282}{\texttt{ arXiv:hep-ph/9903282}}.

\bibitem{Martin:2001es}
\hrefCMSnoop {} {A.~D. Martin, R.~G. Roberts, W.~J. Stirling{ et~al.},
  ``MRST2001: Partons and $\alpha_s$ from precise deep inelastic scattering and
  Tevatron jet data'',} \textit{ Eur. Phys. J. C} \textbf{ 23} (2002) 73--87,
  \href{http://dx.doi.org/10.1007/s100520100842}{\doi{10.1007/s100520100842}},
\href{http://www.arXiv.org/abs/hep-ph/0110215}{\texttt{ arXiv:hep-ph/0110215}}.

\bibitem{ct10}
\hrefCMSnoop {} {H.-L. Lai, M.~Guzzi, J.~Huston{ et~al.}, ``{New parton
  distributions for collider physics}'',} \textit{ Phys. Rev. D} \textbf{ 82}
  (2010) 074024,
  \href{http://dx.doi.org/10.1103/PhysRevD.82.074024}{\doi{10.1103/PhysRevD.82.074024}},
\href{http://www.arXiv.org/abs/1007.2241}{\texttt{ arXiv:1007.2241}}.

\bibitem{P0}
\hrefCMSnoop {} {P.~Skands, ``{The Perugia Tunes}'',} (2009).
\href{http://www.arXiv.org/abs/0905.3418}{\texttt{ arXiv:0905.3418}}.

\bibitem{Jung:2004gs}
\hrefCMSnoop {} {H.~Jung, ``{Un-integrated PDFs in CCFM}'',} (2004).
\href{http://www.arXiv.org/abs/hep-ph/0411287}{\texttt{ arXiv:hep-ph/0411287}}.

\bibitem{Martin:2009iq}
\hrefCMSnoop {} {A.~Martin, W.~Stirling, R.~Thorne{ et~al.}, ``{Parton
  distributions for the LHC}'',} \textit{ Eur. Phys. J. C} \textbf{ 63} (2009)
  189--285,
  \href{http://dx.doi.org/10.1140/epjc/s10052-009-1072-5}{\doi{10.1140/epjc/s10052-009-1072-5}},
  \href{http://www.arXiv.org/abs/0901.0002}{\texttt{ arXiv:0901.0002}}.

\bibitem{Abazov:2008hu}
\hrefCMSnoop {} {{ D0} Collaboration, ``Measurement of the inclusive jet cross
  section in $p \bar{p}$ collisions at $\sqrt{s}=1.96 {\rm TeV}$'',} \textit{
  Phys. Rev. Lett.} \textbf{ 101} (2008) 062001,
  \href{http://dx.doi.org/10.1103/PhysRevLett.101.062001}{\doi{10.1103/PhysRevLett.101.062001}},
\href{http://www.arXiv.org/abs/0802.2400}{\texttt{ arXiv:0802.2400}}.

\bibitem{pdf4lhc}
\hrefCMSnoop {} {S.~Alekhin {et~al.}, ``{The PDF4LHC Working Group Interim
  Report}'',} (2011). \href{http://www.arXiv.org/abs/1101.0536}{\texttt{
  arXiv:1101.0536}}.

\bibitem{Ball:2011mu}
\hrefCMSnoop {} {R.~D. Ball {et~al.}, ``{Impact of Heavy Quark Masses on Parton
  Distributions and LHC Phenomenology}'',} \textit{ Nucl. Phys. B} \textbf{
  849} (2011) 296--363,
  \href{http://dx.doi.org/10.1016/j.nuclphysb.2011.03.021}{\doi{10.1016/j.nuclphysb.2011.03.021}},
\href{http://www.arXiv.org/abs/1101.1300}{\texttt{ arXiv:1101.1300}}.

\bibitem{point6}
\hrefCMSnoop {} {A.~Banfi, G.~P. Salam, and G.~Zanderighi, ``{Phenomenology of
  event shapes at hadron colliders}'',} \textit{ JHEP} \textbf{ 06} (2010) 038,
  \href{http://dx.doi.org/10.1007/JHEP06(2010)038}{\doi{10.1007/JHEP06(2010)038}},
\href{http://www.arXiv.org/abs/1001.4082}{\texttt{ arXiv:1001.4082}}.

\bibitem{Amsler:2008zzb}
\hrefCMSnoop {} {{ Particle Data Group} Collaboration, ``Review of particle
  physics'',} \textit{ Phys. Lett. B} \textbf{ 667} (2008) 101,
\href{http://dx.doi.org/10.1016/j.physletb.2008.07.018}{\doi{10.1016/j.physletb.2008.07.018}}.

\bibitem{Lafferty:1994cj}
\hrefCMSnoop {} {G.~D. Lafferty and T.~R. Wyatt, ``{Where to stick your data
  points: The treatment of measurements within wide bins}'',} \textit{ Nucl.
  Instrum. Meth. A} \textbf{ 355} (1995) 541--547,
\href{http://dx.doi.org/10.1016/0168-9002(94)01112-5}{\doi{10.1016/0168-9002(94)01112-5}}.

\bibitem{Rabbertz:1368241}
\href {http://cdsweb.cern.ch/record/1368241} {{ CMS} Collaboration,
  ``Comparison of Inclusive Jet and Dijet Mass Cross Sections at $\sqrt{s}$ = 7
  TeV with Predictions of perturbative QCD'',} CMS-NOTE 2011-004, CERN, Geneva,
  (2011).

\end{thebibliography}\endgroup

\cleardoublepage \appendix\section{The CMS Collaboration \label{app:collab}}\begin{sloppypar}\hyphenpenalty=5000\widowpenalty=500\clubpenalty=5000\textbf{Yerevan Physics Institute,  Yerevan,  Armenia}\\*[0pt]
S.~Chatrchyan, V.~Khachatryan, A.M.~Sirunyan, A.~Tumasyan
\vskip\cmsinstskip
\textbf{Institut f\"{u}r Hochenergiephysik der OeAW,  Wien,  Austria}\\*[0pt]
W.~Adam, T.~Bergauer, M.~Dragicevic, J.~Er\"{o}, C.~Fabjan, M.~Friedl, R.~Fr\"{u}hwirth, V.M.~Ghete, J.~Hammer\cmsAuthorMark{1}, M.~Hoch, N.~H\"{o}rmann, J.~Hrubec, M.~Jeitler, W.~Kiesenhofer, A.~Knapitsch, M.~Krammer, D.~Liko, I.~Mikulec, M.~Pernicka$^{\textrm{\dag}}$, B.~Rahbaran, C.~Rohringer, H.~Rohringer, R.~Sch\"{o}fbeck, J.~Strauss, A.~Taurok, F.~Teischinger, P.~Wagner, W.~Waltenberger, G.~Walzel, E.~Widl, C.-E.~Wulz
% ...
\vskip\cmsinstskip
\dag:~Deceased\\
1:~~Also at CERN, European Organization for Nuclear Research, Geneva, Switzerland\\
% ...

\end{sloppypar}
\end{document}